\documentclass[12pt]{article}
\usepackage{epsfig}
\usepackage{amsfonts}
\usepackage{latexsym}
\usepackage{amsmath}
\usepackage{mathrsfs}
\usepackage{hyperref}
\numberwithin{equation}{section}

\DeclareFontFamily{OT1}{rsfs}{}
\DeclareFontShape{OT1}{rsfs}{m}{n}{
<-7> rsfs5 <7-10> rsfs7 <10-> rsfs10}{}
\DeclareMathAlphabet{\mycal}{OT1}{rsfs}{m}{n}

\newcommand{\bea}{\begin{eqnarray}}
\newcommand{\eea}{\end{eqnarray}}
\newcommand{\be}{\begin{equation}}
\newcommand{\ee}{\end{equation}}

  \makeatletter
  \let\over=\@@over \let\overwithdelims=\@@overwithdelims
  \let\atop=\@@atop \let\atopwithdelims=\@@atopwithdelims
  \let\above=\@@above \let\abovewithdelims=\@@abovewithdelims
  \makeatother

\begin{document}

\begin{titlepage}
\unitlength = 1mm
\begin{center}

{ \LARGE {\textsc{Towards the quantization of the non-relativistic D2-brane
in the \\ \vskip4mm Pure Spinor Formalism}}}

\qquad

\vspace{0.8cm}

A. Herrera--Aguilar$^{a,b,}$\footnote{Email: aherrera@ifuap.buap.mx},
J.E. Paschalis $^{c,}$\footnote{Email: paschalis@physics.auth.gr}

\vskip7mm

$^a$Instituto de F\'{\i}sica, Benem\'{e}rita Universidad Aut\'{o}noma de Puebla,\\
Apartado Postal J-48, 72570, Puebla, Puebla, M\'exico

$^b$Instituto de F\'{\i}sica y Matem\'aticas, Universidad Michoacana de San Nicol\'{a}s de Hidalgo, \\
Edificio C-$3$,Ciudad Universitaria, CP 58040, Morelia, Michoac\'{a}n, M\'{e}xico\

$^c$ Theoretical Physics Department, Aristotle University of Thessaloniki, \\
54124 Thessaloniki, Greece

\vspace{1.0 cm}
\begin{abstract}
An attempt is made to apply the pure spinor formalism to the non-relativistic IIA D2-brane. The fermionic constraints corresponding to the rescaled fermionic coordinates are given. Two commuting spinor fields are introduced, each one corresponding to a fermionic
constraint. A BRST charge is constructed via the ansatz proposed by N. Berkovits in \cite{NB}--\cite{NBictp}. The nilpotency of 
the BRST charge leads to a set of constraints for the two spinor fields including  pure spinor constraints. A novel non-trivial 
solution is given for one of the spinor fields which can be written as a sum of two pure spinors.
\end{abstract}

\vspace{1.0cm}

\end{center}
\end{titlepage}

\pagestyle{plain}
\setcounter{page}{1}
\newcounter{bean}
\baselineskip18pt


\setcounter{tocdepth}{2}
\section{Introduction}

The pure spinor formalism introduced by N. Berkovits \cite{NB}--\cite{NBictp} is a successful attempt to solve the longstanding problem of finding a manifestly supersymmetric and covariant superstring formalism. The basic ingredient is the BRST-like operator $Q=\int dz \lambda ^{\alpha}d_{\alpha}$ where $d_{\alpha}$ is the fermionic constraint that appears in the conventional Green-Schwarz formalism and $\lambda_{\alpha}$ is a bosonic chiral spinor that plays the role of the associated ``ghost". For $Q$ to be regarded as a BRST operator must be nilpotent and this leads to the relation $\lambda^{\alpha}\gamma_{\alpha \beta}^{m}\lambda^{\beta}=0$. This in 10 dimensions is the condition for $\lambda^{\alpha}$ to be characterized as a pure spinor.

An important property of $Q$ is that its cohomology correctly reproduces the spectrum of the superstring. The pure spinor formalism has been used as well for the covariant quantization of the superparticle \cite{NB2} and also to study several aspects of string theory, for example the propagation of strings in curved backgrounds \cite{CUR}--\cite{CUR3}.

Another important application of the pure spinor formalism is the calculation of scattering amplitudes within the framework of superstring theory \cite{SCA}--\cite{SCA4}. The manifest Lorentz covariance and spacetime supersymmetry make the calculation much easier than in other formalisms. Thus, pure spinors play a crucial role within string perturbation theory. However, within the context of D-branes, there are no non-trivial solutions reported in the literature. Moreover, so far it is not known whether this formalism can be consistently extended or generalized to be applied to non-relativistic systems with kappa symmetry, a fact that could allow us to draw some conclusions for the general case too, especially if there are problems in solving the pure spinor constraints in that case.

In this paper we try to extend the pure spinor formalism to the case of the non-relativistic IIA D2-brane. The non-relativistic
limit of string theories \cite{STR}--\cite{STR1} give us a deeper understanding of string theories themselves. The non-relativistic limit of Dp branes has been studied in \cite{GPR}--\cite{KR}. It is important to note that in this limit the kappa symmetry is maintained and this allows us to treat non-relativistic Dp branes in the framework of the pure spinor formalism.

Here we present a novel non-trivial solution for the non-relativistic D2-brane within the pure spinor formalism. This fact could lead to the quantization of branes with interesting and relevant results.

Our starting point is the action of a IIA D2-brane in a flat 10d background. The fields consist of the 10d superspace coordinates ($x^m,\theta$) and an Abelian gauge field $A_\mu$ \cite{APS}--\cite{HK}:
\begin{equation}
S=-T\int d^{3}\sigma
\sqrt{-\det(G_{\mu\nu}+{\cal F}_{\mu\nu}})+\int {\cal L}_{3}^{WZ},
\label{S}
\end{equation}
where $T$ is the string tension and 
the Wess-Zumino action reads
\begin{equation}
\label{ch1 eq2}
{\cal L}_{3}^{WZ}=T\{\frac{1}{2}d\bar{\theta}\rlap/\Pi^{2}\theta-\frac{1}{3}d\overline{\theta}\rlap/\Pi\theta_{1}+\frac{1}{15}d\overline{\theta}\theta_{2}+d\overline{\theta}\Gamma_{11}\theta {\cal F}\};
\end{equation}
$
G_{\mu\nu}=\eta_{mn}\Pi^{m}_{\mu}\Pi^{n}_{\nu}\
$
and here
$
\ \Pi^{m}_{\mu}=\frac{\partial X^{m}}{\partial
\sigma^{\mu}}-\overline{\theta}\Gamma^{m}\frac{\partial \theta
}{\partial \sigma^{\mu}}.
$
\begin{equation}
{\cal F}_{\mu\nu}=\frac{\partial A_{\nu}}{\partial \sigma^{\mu}}-\frac{\partial A_{\mu}}{\partial \sigma^{\nu}}+\left(\overline{\theta}\Gamma_{11}\Gamma_{m}\partial_{\nu}\theta\right)\left(\Pi^{m}_{\mu}+\frac{1}{2}\overline{\theta}\Gamma^{m}\partial_{\mu}\theta\right)-
\left(\overline{\theta}\Gamma_{11}\Gamma_{m}\partial_{\mu}\theta\right)\left(\Pi^{m}_{\nu}+\frac{1}{2}\overline{\theta}\Gamma^{m}\partial_{\nu}\theta\right)
\end{equation}

\begin{equation}{\theta}_{1}=\rlap/V\theta, \qquad\qquad {\theta}_{2}=\rlap/\widetilde{V}\rlap/V\theta\end{equation}
with
\begin{equation}\rlap/V=\{(\overline{\theta}\Gamma^{m}d\theta)+\Gamma_{11}(\overline{\theta}\Gamma_{11}\Gamma^{m}d\theta)\}\Gamma_{m}\end{equation}

\begin{equation}\rlap/\widetilde{V}=\{(\overline{\theta}\Gamma^{m}d\theta)-\Gamma_{11}(\overline{\theta}\Gamma_{11}\Gamma^{m}d\theta)\}\Gamma_{m}\end{equation}

\begin{equation}\rlap/\Pi=(dX^{m}+\overline{\theta}\Gamma^{m}d\theta)\Gamma_{m}\end{equation}

\begin{equation}{\cal F}=\frac{1}{2}{\cal F}_{\mu\nu}d\sigma^{\mu}d\sigma^{\nu}\end{equation}
and $m,n=0,..,9/   \mu,\nu=0,1,2.$
\vspace{0.3cm}

The action (\ref{S}) has a global supersymmetry and also a local supersymmetry (kappa symmetry).

The action for the non-relativistic D2-brane is obtained from (\ref{S}) by doing the following rescaling \cite{GPR}--\cite{KR}:

\begin{equation} X^{\mu}=\omega x^{\mu}\end{equation}

\begin{equation} X^{a}=x^{a}\end{equation}

\begin{equation} T=\frac{1}{\omega}T_{NR}\end{equation}

\begin{equation} A_{i}=\omega w_{i}\end{equation}
where the subindex $_{NR}$ stands for non-relativistic and $i,\mu=0,1,2/a=3,..,9.$

\begin{equation}\theta=\sqrt{\omega}\theta_{-}+\frac{1}{\sqrt{\omega}}\theta_{+}\end{equation}
where $\theta_{+},\theta_{-}$ are eigenstates of
\begin{equation} \Gamma_{\star}\equiv\Gamma_{0}\Gamma_{1}\Gamma_{2}\end{equation}
\begin{equation}\Gamma_{\star}\theta_{\pm}=\pm \theta_{\pm}\end{equation}

The action of the non-relativistic D2-brane is obtained by expanding (\ref{S}) in powers of $\omega$ and keeping the finite part as
$\omega \rightarrow \infty$ \cite{KR}
\begin{equation}
S_{NR}=\int d^{3}\sigma {\cal L}^{DBI}_{NR}+\int {\cal L}^{WZ}_{NR}
\label{SNR}
\end{equation}
where
\begin{eqnarray}
{\cal L}^{DBI}_{NR}=T_{NR}(\epsilon_{ijk}R^{0}_{i}R^{1}_{j}R^{2}_{k})\left[-\left(\overline{\theta}_{+}\widehat{\gamma^{0}}\frac{\partial\theta_{+}}{\partial\sigma^{0}}+
\overline{\theta}_{+}\widehat{\gamma^{1}}\frac{\partial\theta_{+}}{\partial\sigma^{1}}+\overline{\theta}_{+}\widehat{\gamma^{2}}\frac{\partial\theta_{+}}{\partial\sigma{2}}\right)
+\right. \nonumber \\
\left.
\frac{1}{2}\widehat{g}\,^{il}(\eta_{a{a\acute{}}}\,u^{a}_{i}u^{a\acute{}}_{l})+\frac{1}{4}\widetilde{{\cal F}}^{(1)}_{il}
\widetilde{{\cal F}}^{(1)}_{jk}\widehat{g}\,^{ij}\widehat{g}\,^{lk}\right]\end{eqnarray}

\begin{eqnarray}
{\cal L}^{WZ}_{NR}=
T_{NR}\left[\frac{1}{2}(\overline{\theta}_{+}\Gamma_{\mu\nu}d\theta_{+})
\right.
\Big((dx^{\mu}+\overline{\theta}_{-}\Gamma^{\mu}d\theta_{-})(dx^{\nu}+\overline{\theta}_{-}\Gamma^{\nu}d\theta_{-})
\Big.
\nonumber \\
\Big.
-(\overline{\theta}_{-}\Gamma^{\mu}d\theta_{-})
(dx^{\nu}+\frac{2}{3}\overline{\theta}_{-}\Gamma^{\nu}d\theta_{-})\Big)+ \frac{1}{2}(\overline{\theta}_{-}\Gamma_{\mu\nu}d\theta_{-})
(\overline{\theta}_{+}\Gamma^{\mu}d\theta_{+})\left(dx^{\nu}
+\frac{2}{3}\overline{\theta}_{-}\Gamma^{\nu}d\theta_{-}\right)\nonumber \\ +\frac{1}{2}(\overline{\theta}_{-}\Gamma_{ab}d\theta_{-})\Big(\left(dx^{a}+\overline{\theta}_{+}\Gamma^{a}d\theta_{-}+\overline{\theta}_{-}\Gamma^{a}d\theta_{+}\right)
(dx^{b}+\overline{\theta}_{+}\Gamma^{b}d\theta_{-}+\overline{\theta}_{-}\Gamma^{b}d\theta_{+})
\Big.
\nonumber \\
\left.
-(\overline{\theta}_{+}\Gamma^{a}d\theta_{-}+
+\overline{\theta}_{-}\Gamma^{a}d\theta_{+})\left(dx^{b}+\frac{2}{3}(\overline{\theta}_{-}\Gamma^{b}d\theta_{+}+\overline{\theta}_{+}\Gamma^{b}d\theta_{-})\right)\right)+\nonumber\\
(\overline{\theta}_{+}\Gamma_{\mu a}d\theta_{-}+\overline{\theta}_{-}\Gamma_{\mu a}d\theta_{+})\left((dx^{\mu}+\overline{\theta}_{-}\Gamma^{\mu}d\theta_{-})\left(dx^{a}+\frac{1}{2}(\overline{\theta}_{+}\Gamma^{a}d\theta_{-}+\overline{\theta}_{-}
\Gamma^{a}d\theta_{+}\right)\right)\nonumber \\
-\frac{1}{2}(\overline{\theta}_{-}\Gamma^{\mu}d\theta_{-})\left(dx^{a}+\frac{1}{3}(\overline{\theta}_{-}\Gamma^{a}d\theta_{+}+\overline{\theta}_{+}\Gamma^{a}d\theta_{-})\right)
+\frac{1}{2}(\overline{\theta}_{+}\Gamma_{\mu}\Gamma_{11}d\theta_{-}+\overline{\theta}_{-}\Gamma_{\mu}\Gamma_{11}d\theta_{+})\nonumber \\
\left(dx^{\mu}+\frac{2}{3}\overline{\theta}_{-}\Gamma^{\mu}d\theta_{-}\right)(\overline{\theta}_{+}\Gamma_{11}d\theta_{-}+\overline{\theta}_{-}\Gamma_{11}d\theta_{+})
+\frac{1}{2}(\overline{\theta}_{-}\Gamma_{b}\Gamma_{11}d\theta_{-})\Big(dx^{b}+\frac{2}{3}(\overline{\theta}_{-}\Gamma^{b}d\theta_{+}+
\Big.
\nonumber \\
\Big.
\overline{\theta}_{+}\Gamma^{b}d\theta_{-})\Big)(\overline{\theta}_{+}\Gamma_{11}d\theta_{-}+\overline{\theta}_{-}\Gamma_{11}d\theta_{+})+
(d\overline{\theta}_{-}\Gamma_{11}\theta_{+}+d\overline{\theta}_{+}\Gamma_{11}\theta_{-})(f-(\overline{\theta}_{-}\Gamma_{\mu}\Gamma_{11}d\theta_{+}+\nonumber \\
\left.
\overline{\theta}_{+}\Gamma_{\mu}\Gamma_{11}d\theta_{-})\left(dx^{\mu}+\frac{1}{2}\overline{\theta}_{-}\Gamma^{\mu}d\theta_{-}\right)-(\overline{\theta}_{-}
\Gamma_{a}\Gamma_{11}d\theta_{-})\left(dx^{a}+\frac{1}{2}(\overline{\theta}_{-}\Gamma^{a}d\theta_{+}+\overline{\theta}_{+}\Gamma^{a}d\theta_{-})\right)\right]
\label{LWZNR}
\end{eqnarray}
where
\begin{eqnarray}R^{\mu}_{i}=\partial_{i}x^{\mu}-\overline{\theta}_{-}\Gamma^{\mu}\partial_{i}\theta_{-}\\
u^{a}_{i}=\partial_{i}x^{a}-\overline{\theta}_{-}\Gamma^{a}\partial_{i}\theta_{+}-\overline{\theta}_{+}\Gamma^{a}\partial_{i}\theta_{-}\end{eqnarray}
$i,\mu=0,1,2/a=3,..,9;$
also\
\begin{eqnarray}\widehat{\gamma^{0}}=\frac{1}{(\epsilon_{ijk}R^{0}_{i}R^{1}_{j}R^{2}_{k})}(\epsilon_{ijk}\Gamma^{i}R^{j}_{1}R^{k}_{2})\\
\widehat{\gamma^{1}}=-\frac{1}{(\epsilon_{ijk}R^{0}_{i}R^{1}_{j}R^{2}_{k})}(\epsilon_{ijk}\Gamma^{i}R^{j}_{0}R^{k}_{2})\\
\widehat{\gamma^{2}}=\frac{1}{(\epsilon_{ijk}R^{0}_{i}R^{1}_{j}R^{2}_{k})}(\epsilon_{ijk}\Gamma^{i}R^{j}_{0}R^{k}_{1})\\
\widehat{g}^{jk}=\eta_{\mu\nu}R^{\mu}_{j}R^{\nu}_{k}\end{eqnarray}
where we have introduced the following quantities $\epsilon_{012}=1,\ $
$\eta_{\mu\nu}=\text{diag}(-1,1,1),\ $
$\eta_{a{a\acute{}}}=\text{diag}(1,..,1),\ $
$\mu,\nu=0,1,2;\ $
$a,a\acute{}=3,..,9;$\
\begin{eqnarray}\widetilde{{\cal F}}^{(1)}_{ij}=f_{ij}+\left[(\overline{\theta}_{-}\Gamma_{11}\Gamma_{\mu}\partial_{j}\theta_{+}+\overline{\theta}_{+}
\Gamma_{11}\Gamma_{\mu}\partial_{j}\theta_{-})\left(R^{\mu}_{i}+\frac{1}{2}\overline{\theta}_{-}\Gamma^{\mu}\partial_{i}\theta_{-}\right)-(i\leftrightarrow j)\right]
\nonumber \\
+\left[(\overline{\theta}_{-}\Gamma_{11}\Gamma_{a}\partial_{j}\theta_{-})\left(u^{a}_{i}+\frac{1}{2}(\overline{\theta}_{-}\Gamma^{a}\partial_{i}\theta_{+} + \overline{\theta}_{+}\Gamma^{a}\partial_{i}\theta_{-})\right)-(i\leftrightarrow
j )\right],
\end{eqnarray}
where $f_{ij}=\partial_{i}w_{j}-\partial_{j}w_{i},\ $ $f=\frac{1}{2}f_{ij}d\sigma^{i}d\sigma^{j}, \ $ and $\mu,\nu,i,j=0,1,2/a=3,..,9.$

We denote by $[{\cal L}^{WZ}_{NR}]_{3}$  the  3-form coefficient of ${\cal L}^{WZ}_{NR}$ given in (\ref{LWZNR}). The Lagrangian density of the non-relativistic D2-brane is given by:

 \begin{equation}
 {\cal L}_{NR}={\cal L}^{DBI}_{NR}+[{\cal L}^{WZ}_{NR}]_{3}
 \label{LNR}
 \end{equation}

 The action (\ref{SNR}) is invariant under the non-relativistic counterpart of the global and local supersymmetric transformations that leave the relativistic D2-brane action (\ref{S}) invariant \cite{GPR}.

 The conjugate momenta of the variables $x^{\mu},x^{a},\theta_{+},\theta_{-},w_{i}$ are given by:
 \begin{eqnarray}
 p_{\mu}=\frac{\partial{\cal L}_{NR}}{\partial{\dot{x}^{\mu}}}=\nonumber \\
 \frac{\partial{\cal L}_{NR}}{\partial R^{\mu}_{0}}+ \frac{\partial{\cal L}_{NR}}{\partial {\widetilde{{\cal F}}^{(1)}_{0i}}}(\overline{\theta}_{-}\Gamma_{11}\Gamma_{\mu}\partial_{i}\theta_{+}+\overline{\theta}_{+}\Gamma_{11}\Gamma_{\mu}\partial_{i}\theta_{-})
 \end{eqnarray}
$\mu=0,1,2/i=1,2;$
 \begin{eqnarray}p_{a}=\frac{\partial{\cal L}_{NR}}{\partial{\dot{x}^{a}}}=\nonumber \\
 \frac{\partial{\cal L}_{NR}}{\partial{\dot{u}^{a}_{0}}}+\frac{\partial{\cal L}_{NR}}{\partial\widetilde{{\cal F}}^{(1)}_{0i}}(\overline{\theta}_{-}\Gamma_{11}\Gamma_{a}\partial_{i}\theta_{-})\end{eqnarray}
$a=3,..,9;$
\begin{eqnarray}J_{+l}=\frac{\partial^{r}{\cal L}_{NR}}{\partial{\dot{\theta}_{+l}}}-\frac{\partial{\cal L}_{NR}}{\partial u^{a}_{0}}(
\overline{\theta}_{-}\Gamma^{a})_{l}-\nonumber \\
 \frac{\partial{\cal L}_{NR}}{\partial\widetilde{{\cal F}}^{(1)}_{0i}}\left[\left(R^{\mu}_{i}+\frac{1}{2}\overline{\theta}_{-}\Gamma^{\mu}\partial_{i}\theta_{-}\right)(\overline{\theta}_{-}\Gamma_{11}\Gamma_{\mu})_{l}+
 \frac{1}{2}(\overline{\theta}_{-}\Gamma_{11}\Gamma_{a}\partial_{i}\theta_{-})(\overline{\theta}_{-}\Gamma^{a})_{l}\right]\end{eqnarray}
 and
\begin{eqnarray}J_{-l}=\frac{\partial^{r}{\cal L}_{NR}}{\partial{\dot{\theta}_{-l}}}=\frac{\partial^{r}[{\cal L}^{WZ}_{NR}]_{3}}
 {\partial{\dot{\theta}_{-l}}}-\frac{\partial{\cal L}_{NR}}{\partial R^{\mu}_{0}}(\overline{\theta}_{-}\Gamma^{\mu})_{l}-
 \nonumber \\
 \frac{\partial{\cal L}_{NR}}{\partial u^{a}_{0}}(\overline{\theta}_{+}\Gamma^{a})_{l}-\frac{{\cal L}_{NR}}{\partial\widetilde{{\cal F}}^{(1)}
 _{0i}}\left[\frac{1}{2}(\overline{\theta}_{-}\Gamma_{11}\Gamma_{\mu}\partial_{i}\theta_{+}+\overline{\theta}_{+}\Gamma_{11}\Gamma_{\mu}\partial_{i}\theta_{-})
 (\overline{\theta}_{-}\Gamma^{\mu})_{l}+
 \right.
 \nonumber \\
 \frac{1}{2}(\overline{\theta}_{-}\Gamma_{11}\Gamma_{a}\partial_{i}\theta_{-})(\overline{\theta}_{+}\Gamma^{a})_{l}+\left(R^{\mu}_{i}+
 \frac{1}{2}\overline{\theta}_{-}\Gamma^{\mu}\partial_{i}\theta_{-}\right)(\overline{\theta}_{+}\Gamma_{11}\Gamma_{\mu})_{l}+\nonumber \\
 \left.
 \left(u^{a}_{i}+\frac{1}{2}(\overline{\theta}_{-}\Gamma^{a}\partial_{i}\theta_{+}+\overline{\theta}_{+}\Gamma^{a}\partial_{i}\theta_{-})\right)
 (\overline{\theta}_{-}\Gamma_{11}\Gamma_{a})_{l}\right],
 \end{eqnarray}
where $l=1,..,32;$ and
 \begin{equation}
 E^{i}=\frac{\partial{\cal
 L}_{NR}}{\partial{\dot{w}_{i}}}=\frac{\partial{\cal L}_{NR}}{\partial\widetilde{{\cal F}}^{(1)}_{0i}}\,.
 \end{equation}

 The fermionic constraints are given by
 \begin{eqnarray}
 F_{+l}=J_{+l}+p_{a}(\overline{\theta}_{-}\Gamma^{a})_{l}+E^{i}\left[\left(R^{\mu}_{i}+\frac{1}{2}\overline{\theta}_{-}\Gamma^{\mu}\partial_{é}
 \theta_{-}\right)(\overline{\theta}_{-}\Gamma_{11}\Gamma_{\mu})_{l}
 \right.
 \nonumber \\
 \left.
 -\frac{1}{2}(\overline{\theta}_{-}\Gamma_{11}\Gamma_{a}\partial_{i}\theta_{-})(\overline{\theta}_{-}\Gamma^{a})_{l}\right]+T_{NR}\epsilon_{ijk}
 (\overline{\theta}_{+}\Gamma^{i})_{l}R^{j}_{1}R^{k}_{2}-\frac{\partial^{r}[{\cal L}^{WZ}_{NR}]_{3}}{\partial{\dot{\theta}_{+l}}}\end{eqnarray}
where the last derivative does not include differentiation with respect to $u^{a}_{0},\widetilde{{\cal F}}^{(1)}_{0i}$, and
 \begin{eqnarray}
 F_{-l}=J_{-l}+p_{a}(\overline{\theta}_{+}\Gamma^{a})_{l}+p_{\mu}(\overline{\theta}_{-}\Gamma^{\mu})_{l}+E^{i}\left[\left(R^{\mu}_{i}
 +\frac{1}{2}(\overline{\theta}_{-}\Gamma^{\mu}\partial_{i}\theta_{-}\right)(\overline{\theta}_{+}\Gamma_{11}\Gamma_{\mu})_{l}+
\right.
\nonumber \\
 \left(u^{a}_{i}+\frac{1}{2}(\overline{\theta}_{-}\Gamma^{a}\partial_{i}\theta_{+}+\overline{\theta}_{+}\Gamma^{a}\partial_{i}\theta_{-}\right)
(\overline{\theta}_{-}\Gamma_{11}\Gamma_{a})_{l}-\frac{1}{2}(\overline{\theta}_{-}\Gamma_{11}\Gamma_{\mu}\partial_{i}\theta_{+}+\nonumber
 \\
 \left.
 \overline{\theta}_{+}\Gamma_{11}\Gamma_{\mu}\partial_{i}\theta_{-})(\overline{\theta}_{-}\Gamma^{\mu})_{l}-\frac{1}{2}(\overline{\theta}_{-}
 \Gamma_{11}\Gamma_{a}\partial_{i}\theta_{-})(\overline{\theta}_{+}\Gamma^{a})_{l}\right]-\frac{\partial^{r}[{\cal L}^{WZ}_{NR}]_{3}}{\partial
 {\dot{\theta}_{-l}}}\,,
 \end{eqnarray}
where the last derivative does not include differentiation with respect to $u^{a}_{0}, \,R^{\mu}_{0}, \,\widetilde{{\cal F}}^{(1)}_{0i}\,.$

 We introduce now two commuting spinor fields
 $\lambda_{+},\lambda_{-}$ corresponding to the fermionic
 constraints $F_{+},F_{-}$ and we write down a BRST charge
 according to the ansatz proposed in \cite{NB}--\cite{NBictp}:

 \begin{eqnarray}
 \label{Q}
 Q=\int
 d^{2}\sigma\Big\{\lambda^{l}_{+}J_{+l}+\lambda^{l}_{-}J_{-l}+p_{a}(\overline{\theta}_{-}\Gamma^{a}\lambda_{+}+\overline{\theta}_{+}\Gamma^{a}\lambda_{-}
 )+p_{\mu}(\overline{\theta}_{-}\Gamma^{\mu}\lambda_{-})+\nonumber \\
 E^{i}\left[\left(R^{\mu}_{i}+\frac{1}{2}\overline{\theta}_{-}\Gamma^{\mu}\partial_{i}\theta_{-}\right)
 (\overline{\theta}_{+}\Gamma_{11}\Gamma_{\mu}\lambda_{-}
 +\overline{\theta}_{-}\Gamma_{11}\Gamma_{\mu}\lambda_{+})+\Big(u^{a}_{i}+\frac{1}{2}(\overline{\theta}_{-}\Gamma^{a}\partial_{i}\theta_{+}+
 \right.\Big.
 \nonumber \\
 \Big.
 \overline{\theta}_{+}\Gamma^{a}\partial_{i}\theta_{-})\Big)(\overline{\theta}_{-}\Gamma_{11}\Gamma_{a}\lambda_{-})-\frac{1}{2}(\overline{\theta}_{-}
 \Gamma_{11}\Gamma_{\mu}\partial_{i}\theta_{+}+\overline{\theta}_{+}\Gamma_{11}\Gamma_{\mu}\partial_{i}\theta_{-})(\overline{\theta}_{-}\Gamma^{\mu}
 \lambda_{-})-\nonumber \\
 \left.
 \frac{1}{2}(\overline{\theta}_{-}\Gamma_{11}\Gamma_{a}\partial_{i}\theta_{-})(\overline{\theta}_{+}\Gamma^{a}\lambda_{-}+\overline{\theta}_{-}
 \Gamma^{a}\lambda_{+})\right]+T_{NR}\epsilon_{ijk}(\overline{\theta}_{+}\Gamma^{i}\lambda_{+})R^{j}_{1}R^{k}_{2}-\left(\lambda^{T}_{+}\frac{\partial^{r}
 [{\cal L}^{WZ}_{NR}]_{3}}{\partial{\dot{\theta}_{+}}}\right)-\nonumber \\
\left(\lambda^{T}_{-}\frac{\partial^{r}[{\cal L}^{WZ}_{NR}]_{3}}{\partial{\dot{\theta}_{-}}}\right)\Big\}
\end{eqnarray}
with $i,j,k,\mu=0,1,2/a=3,..,9.$

 A set of BRST transformations is the following:
 \begin{eqnarray}s\theta^{l}_{+}=\lambda^{l}_{+},\nonumber \\
 s\theta^{l}_{-}=\lambda^{l}_{-},\nonumber \\
 sx^{\mu}=\overline{\theta}_{-}\Gamma^{\mu}\lambda_{-},\nonumber\\
 sx^{a}=\overline{\theta}_{-}\Gamma^{a}\lambda_{+}+\overline{\theta}_{+}\Gamma^{a}\lambda_{-},\nonumber \\
 sR^{\mu}_{\nu}=2(\partial_{\nu}\overline{\theta}_{-}\Gamma^{\mu}\lambda_{-}),\nonumber \\
 su^{a}_{\mu}=2(\partial_{\mu}\overline{\theta}_{-}\Gamma^{a}\lambda_{+}+\partial_{\mu}\overline{\theta}_{+}\Gamma^{a}\lambda_{-}),\nonumber\\
 s\widetilde{{\cal F}}^{(1)}_{ij}=2R^{\mu}_{j}(\overline{\lambda}_{-}\Gamma_{11}\Gamma_{\mu}\partial_{i}\theta_{+}+\overline{\lambda}+
 \Gamma_{11}\Gamma_{\mu}\partial_{i}\theta_{-})+2u^{a}_{j}(\partial_{i}\overline{\theta}_{-}\Gamma_{11}\Gamma^{a}\lambda_{-})-(i\leftrightarrow j
  ) \end{eqnarray}
where again $\mu,\nu=0,1,2/i,j=1,2/a=3,..,9/l=1,..32;\ $ from which we obtain

  \begin{eqnarray}s^{2}\theta^{l}_{+}=s^{2}\theta^{l}_{-}=0,\nonumber \\
   s^{2}x^{\mu}=\overline{\lambda}_{-}\Gamma^{\mu}\lambda_{-},\nonumber \\
  s^{2}x^{a}=2(\overline{\lambda}_{-}\Gamma^{a}\lambda_{+}),\nonumber \\
  s^{2}R^{\mu}_{i}=\frac{\partial}{\partial\sigma^{i}}(\overline{\lambda}_{-}\Gamma^{\mu}\lambda_{-}),\nonumber \\
  s^{2}u^{a}_{i}=2\frac{\partial}{\partial\sigma^{i}}(\overline{\lambda}_{+}\Gamma^{a}\lambda_{-})\end{eqnarray}
and
\begin{eqnarray}s^{2}\widetilde{{\cal F}}^{(1)}_{ij}=2R^{\mu}_{j}\frac{\partial}{\partial\sigma^{i}}(\overline{\lambda}_{+}\Gamma_{11}\Gamma_{\mu}\lambda_{-})+u^{a}_{j}\frac{\partial}{\partial\sigma^{i}}
  (\overline{\lambda}_{-}\Gamma_{11}\Gamma_{a}\lambda_{-})+\nonumber \\
  2(\partial_{j}\overline{\theta}_{-}\Gamma_{11}\Gamma^{\mu}\partial_{i}\theta_{-})(\overline{\lambda}_{-}\Gamma_{\mu}\lambda_{-})
  +2(\partial_{j}\overline{\theta}_{-}\Gamma^{a}\partial_{i}\theta_{+})(\overline{\lambda}_{-}\Gamma_{11}\Gamma_{a}\lambda_{-})+\nonumber \\
  2(\partial_{j}\overline{\theta}_{-}\Gamma_{11}\Gamma_{a}\partial_{i}\theta_{-})(\overline{\lambda}_{-}\Gamma^{a}\lambda_{+})+2(\partial_{j}\theta_{-}
  \Gamma^{\mu}\partial_{i}\theta_{-})(\overline{\lambda}_{-}\Gamma_{11}\Gamma_{\mu}\lambda_{+})\end{eqnarray}
where we used the ansatz
  \begin{eqnarray}
  \label{P_}
  \lambda_{-}=P_{-}\lambda,\nonumber\\
  P_{-}=\frac{1}{2}(1-\Gamma_{0}\Gamma_{1}\Gamma_{2})
  \end{eqnarray}

The full set of constraints required for the nilpotency of the BRST charge is obtained by studying the equal time Poisson bracket $\{Q(\sigma),Q(\sigma')\}$ where $Q(\sigma)$ is given in (\ref{Q}). A basis for expanding a $32\times 32$ matrix M is given in Appendix A. 

Thus, the  expression for the equal time Poisson bracket $\{Q(\sigma),Q(\sigma')\}$ where $Q(\sigma)$ is the full BRST charge defined in (\ref{Q}) is given by:

  \begin{eqnarray}
  \{Q(\sigma),Q(\sigma')\}=\int d^{2}\sigma(\partial_{i}E^{i})[(\overline{\theta}_{-}\Gamma^{a}\lambda_{+}+\overline{\theta}_{+}\Gamma^{a}\lambda_{-})(\overline{\theta}_{-}\Gamma_{11}\Gamma_{a}
  \lambda_{-})+\nonumber \\
  (\overline{\theta}_{-}\Gamma^{\mu}\lambda_{-})(\overline{\theta}_{+}\Gamma_{11}\Gamma_{\mu}\lambda_{-}+\overline{\theta}_{-}\Gamma_{11}\Gamma_{\mu}\lambda_{+})]+\nonumber \\
  2T_{NR}\epsilon_{\mu\nu\rho}\int d^{2}\sigma R^{\nu}_{1}R^{\rho}_{2}(\overline{\lambda}_{+}\Gamma^{\mu}\lambda_{+})+T_{NR}\epsilon_{0jk}\int d\sigma^{2}R^{\mu}_{j}R^{\nu}_{k}(\overline{\lambda}_{+}\Gamma_{\mu\nu}\lambda_{+})+\nonumber \\
  \int d^{2}\sigma\Big[F_{\mu}(\overline{\lambda}_{-}\Gamma^{\mu}\lambda_{-})+F_{a}(\overline{\lambda}_{-}\Gamma^{a}\Gamma_{11}\lambda_{-})+F_{ab}(
  \overline{\lambda}_{-}\Gamma^{ab}\lambda_{-})+\nonumber \\
  F^{\nu abdf}(\overline{\lambda}_{-}\Gamma_{\nu abdf}\lambda_{-})+\overline{F}(\overline{\lambda}_{-}\Gamma_{11}\lambda_{+})+
  \overline{F}_{a}(\overline{\lambda}_{-}\Gamma^{a}\lambda_{+})+\nonumber \\
  \overline{F}_{\nu b}(\overline{\lambda}_{-}\Gamma^{\nu b}\lambda_{+})+\overline{F}_{\nu}(\overline{\lambda}_{-}\Gamma^{\nu}\Gamma_{11}\lambda_{+})+
  \overline{F}_{\nu abc}(\overline{\lambda}_{-}\Gamma^{\nu
  abc}\lambda_{+})+\overline{F}^{\nu ab}(\overline{\lambda}_{+}\Gamma_{\nu ab}\Gamma_{11}\lambda_{-})+\nonumber \\
  \overline{F}_{fglmn}(\overline{\lambda}_{+}\Gamma^{fglmn}\lambda_{-})+\overline{F}_{lmn}(\overline{\lambda}_{-}\Gamma^{lmn}\lambda_{+})\Big]  \end{eqnarray}
where now $\epsilon_{012}=1,$ $i=1,2/\mu,\nu,\rho=0,1,2/a,b,c,d,f,g,l,m,n=3,..,9;\ $ and the ansatz $\lambda_{-}=P_{-}\lambda$ was used.

The expressions for $ F_{\mu},F_{a},F_{ab},F_{\nu abdf},\overline{F},\overline{F}_{a},
\overline{F}_{\nu},\overline{F}_{\nu b },\overline{F}_{\nu abc},\overline{F}_{\nu ab},\overline{F}_{fglm},\overline{F}_{lmn}$
are given in Appendix B. These expressions are different from zero, so the following constraints guarantee the nilpotency of the BRST
charge:

\begin{eqnarray}
\label{Gauss}
\partial_{i}E^{i}=0 \qquad (\text{Gauss law}),\nonumber \\
\overline{\lambda}_{+}\Gamma_{\mu}\lambda_{+}=0, \qquad
\overline{\lambda}_{+}\Gamma_{\mu\nu}\lambda_{+}=0,\nonumber \\
\overline{\lambda}_{-}\Gamma_{\mu}\lambda_{-}=0,\qquad
\overline{\lambda}_{-}\Gamma_{a}\Gamma_{11}\lambda_{-}=0,\qquad
\overline{\lambda}_{-}\Gamma_{ab}\lambda_{-}=0, \qquad
\overline{\lambda}_{-}\Gamma_{\mu abcd}\lambda_{-}=0,\nonumber \\
\overline{\lambda}_{-}\Gamma_{11}\lambda_{+}=0,\qquad
\overline{\lambda}_{-}\Gamma_{a}\lambda_{+}=0,\qquad
\overline{\lambda}_{-}\Gamma_{\mu a}\lambda_{+}=0,\qquad
\overline{\lambda}_{-}\Gamma_{\mu}\Gamma_{11}\lambda_{+}=0,\nonumber
\\
\overline{\lambda}_{-}\Gamma_{\mu abc}\lambda_{+}=0, \qquad
\overline{\lambda}_{+}\Gamma_{\mu ab}\Gamma_{11}\lambda_{-}=0, \qquad
\overline{\lambda}_{+}\Gamma_{fglmn}\lambda_{-}=0, \qquad
\overline{\lambda}_{-}\Gamma_{lmn}\lambda_{+}=0
\end{eqnarray}
with the ansatz $\lambda_{-}=P_{-}\lambda,(
i=1,2/\mu,\nu=0,1,2/a,b,c,d,f,g,l,m,n=3,..,9)\,.$
%
%

\section{Solving the constraints}

In order to solve the constraints (\ref{Gauss}) we use the methodology
used in \cite{GPN}--\cite{GPN1}. The Dirac matrices $ \Gamma^{m}\, (m=0,1,..,9)$
are combined into five creation operators $a^{i}\, (i=1,..,5)$ and
five annihilation operators $a_{i}\, (i=1,..,5)$ as follows:

\begin{eqnarray}
a^{1}=\frac{1}{2}(\Gamma^{1}-i\Gamma^{2}),\qquad
a^{2}=\frac{1}{2}(\Gamma^{3}-i\Gamma^{4}),\qquad
a^{3}=\frac{1}{2}(\Gamma^{5}-i\Gamma^{6}),\nonumber \\
a^{4}=\frac{1}{2}(\Gamma^{7}-i\Gamma^{8}),\qquad
a^{5}=\frac{1}{2}(\Gamma^{9}+\Gamma^{0}),\nonumber \\
a_{1}=\frac{1}{2}(\Gamma^{1}+i\Gamma^{2}),\qquad
a_{2}=\frac{1}{2}(\Gamma^{3}+i\Gamma^{4}),\qquad
a_{3}=\frac{1}{2}(\Gamma^{5}+i\Gamma^{6}),\nonumber \\
a_{4}=\frac{1}{2}(\Gamma^{7}+i\Gamma^{8}),\qquad
a_{5}=\frac{1}{2}(\Gamma^{9}-\Gamma^{0})
\end{eqnarray}

The following identities hold:
\begin{eqnarray}\{a_{i},a^{j}\}=\delta_{ij},\{a^{i},a^{j}\}=0,\qquad
\{\Gamma_{11},a^{i}\}=0,\qquad
\{\Gamma_{11},a_{j}\}=0,\qquad
a_{i}=a^{i\dag},\nonumber \\ \text{with now}\quad
i,j=1,..,5.
\end{eqnarray}

Here we introduce the vacuum state  $\mid0>$ and the state $<0\mid$ with the equations:
\begin{eqnarray} a_{i}\mid0>=0, \qquad
<0\mid a^{i}=0, \qquad i=1,..,5. \end{eqnarray}
We decompose a 32-component Dirac spinor into a sum of a positive
chirality 16-component part and a negative chirality 16-component
part as follows:
\begin{eqnarray}\mid \lambda> = \left(\lambda_{+}\mid 0>+
\frac{1}{2}\lambda_{ij}a^{j}a^{i}\mid 0>+\frac{1}{4!}\lambda^{i}\epsilon_{ijklm}a^{j}a^{k}a^{l}a^{m}
\mid 0>\right)+\nonumber \\
\left(\frac{1}{5!}\lambda'_{+}\epsilon_{ijklm}a^{i}a^{j}a^{k}a^{l}a^{m}\mid
0> +\frac{1}{3!}\lambda'^{ij}\epsilon_{ijklm}a^{k}a^{l}a^{m}\mid
0>+
\lambda'_{i}a^{i}\mid 0>\right)=
\label{spinor}
\\
\nonumber
(1+\overline{10}+5)+(1+10+\overline{5}),
\end{eqnarray}
where $\lambda_{ij}=-\lambda_{ji},\lambda'^{ij}=-\lambda'^{ji}\,.$

Let us try to solve from (\ref{Gauss}) the constraints $\overline{\lambda}_{+}\Gamma_{\mu}\lambda_{+}=0$ and
$\overline{\lambda}_{+}\Gamma_{\mu\nu}\lambda_{+}=0\,$ (where $\mu,\nu=0,1,2$).

It is interesting to note that $\lambda_{+}$ can be written as a sum of two pure spinors $ \lambda_{1+}=P_{+}\lambda_{+}$ and $\lambda_{2+}=P_{-}\lambda_{+}$ where $P_{\pm}=\frac{1}{2}(1\pm
\Gamma_{0}\Gamma_{1}\Gamma_{2})$ and the constraints
$\overline{\lambda}_{+}\Gamma_{\mu}\lambda_{+}=0$ ,
$\overline{\lambda}_{+}\Gamma_{\mu\nu} \lambda_{+}=0,
(\mu,\nu=0,1,2)$ can be expressed equivalently as two pure spinor constraints: $\overline{\lambda}_{1+}\Gamma_{m}\lambda_{1+}=0$, $\overline{\lambda}_{2+}\Gamma_{m} \lambda_{2+}=0, (m=0,...,9)$.\\

The constraints $\overline{\lambda}_{+}\Gamma_{\mu}\lambda_{+}=0$ can be written as follows:
\begin{eqnarray}
\label{Ca5}
<\lambda_{+}\mid Ca_{1}\mid \lambda_{+}>=0,\nonumber \\
<\lambda_{+}\mid Ca^{1}\mid \lambda_{+}>=0,\nonumber \\
<\lambda_{+}\mid Ca_{5}\mid \lambda_{+}>=<\lambda_{+}\mid Ca^{5}\mid \lambda_{+}>\,,
\end{eqnarray}

%
%
%
%
%
whereas the constraints
$\overline{\lambda}_{+}\Gamma_{\mu\nu}\lambda_{+}=0, (\mu,\nu=0,1,2)$
can be expressed as:
\begin{eqnarray}<\lambda_{+}\mid C(a^{5}-a_{5})(a^{1}+a_{1})\mid
\lambda_{+}>=0,\nonumber \\
<\lambda_{+}\mid C(a^{1}+a_{1})(a^{1}-a_{1})\mid
\lambda_{+}>=0,\nonumber \\
<\lambda_{+}\mid C(a^{5}-a_{5})(a^{1}-a_{1})\mid
\lambda_{+}>=0\,.
\end{eqnarray}



For a general $\mid \lambda_{+}>$ we can write
\begin{eqnarray} \mid \lambda_{+}>=\lambda_{++}\mid
0>+\frac{1}{2}\lambda_{+ij}a^{j}a^{i}\mid
0>+\frac{1}{4!}\lambda^{i}_{+}\epsilon_{ijklm}a^{j}a^{k}a^{l}a^{m}\mid
0>+\nonumber \\
\lambda'_{+i}a^{i}\mid
0>+\frac{1}{3!}\lambda'^{ij}_{+}\epsilon_{ijlkm}a^{l}a^{k}a^{m}\mid
0>+\frac{1}{5!}\lambda'_{++}\epsilon_{ijklm}a^{i}a^{j}a^{k}a^{l}a^{m}\mid
0>,
\end{eqnarray}
where now $\epsilon_{12345}=1.$

Thus we have the following relations:
\begin{equation}
\label{211}
<\lambda_{+}\mid Ca^{1}\mid
\lambda_{+}>=2\left(\lambda_{++}\lambda^{1}_{+}+\frac{1}{8}\epsilon_{1ijkl}\lambda_{+ij}\lambda_{+lk}\right)
-4\lambda'_{+i}\lambda^{'i1}_{+}=0,
\end{equation}
\begin{equation}
\label{212}
<\lambda_{+}\mid Ca_{1}\mid \lambda_{+}>=
2\lambda^{i}\lambda_{+i1}-2\lambda'_{+1}\lambda'_{++}+\epsilon_{1iji'j'}\lambda^{'ij}_{+}
\lambda^{'i'j'}_{+}=0,
\end{equation}
\begin{eqnarray}
\label{213}
<\lambda_{+}\mid Ca^{5}\mid
\lambda_{+}>=<\lambda_{+}\mid Ca_{5}\mid \lambda_{+}> \qquad\qquad\Rightarrow
\nonumber \\
2(\lambda_{++}\lambda^{5}_{+}+\frac{1}{8}\epsilon_{5ijkl}\lambda_{+ij}\lambda_{+lk})-4\lambda'_{+i}\lambda^{'i5}_{+}=
2\lambda^{i}_{+}\lambda_{+i5}-2\lambda'_{+5}\lambda'_{++}+\epsilon_{5iji'j'}\lambda^{'ij}_{+}\lambda^{'i'j'}_{+}.
\end{eqnarray}

On the other hand the constraints $\overline{\lambda}_{+}\Gamma^{\mu\nu}\lambda_{+}=0$ can be rewritten as follows:
\begin{eqnarray}
\label{214}
\frac{1}{2}<\lambda_{+}\mid
C(a_{1}a^{1}-a^{1}a_{1})\mid \lambda_{+}>=\nonumber \\
-\lambda_{++}\lambda'_{++}+4\lambda_{+i1}\lambda^{'i1}_{+}-\lambda_{+ij}\lambda^{'ij}+\lambda^{i}_{+}\lambda'_{+i}-2\lambda^{1}_{+}
\lambda'_{+1}=0,\nonumber \\
<\lambda_{+}\mid C(a^{5}-a_{5})a^{1}\mid \lambda_{+}>=\nonumber \\
4\lambda_{++}\lambda^{'51}_{+}+4\lambda^{'i1}_{+}\lambda_{+i5}+2\lambda'_{+5}\lambda^{1}_{+}-\epsilon_{15iji'}\lambda_{+ij}\lambda'_{+i'}=0,\nonumber\\
\frac{1}{2}<\lambda_{+}\mid C(a^{5}-a_{5})a_{1}\mid \lambda_{+}>=\nonumber \\
-2\lambda_{+i1}\lambda^{'i5}_{+}+\lambda_{+51}\lambda'_{++}+\lambda^{5}_{+}\lambda'_{+1}-\epsilon_{ijk15}\lambda^{'ij}_{+}\lambda^{k}
_{+}=0.
\end{eqnarray}

From (\ref{211})-(\ref{214}) we obtain the non-trivial solution
\begin{eqnarray} 
\label{soln}
\lambda^{1}_{+}=\frac{1}{\lambda}_{++}(2\lambda'_{+i}\lambda^{'i1}_{+}+\frac{1}{8}\epsilon_{1ijkl}\lambda_{+ij}\lambda_{+kl}),\nonumber \\
\lambda'_{+1}=\frac{1}{2\lambda'_{++}}(2\lambda^{i}_{+}\lambda_{+i1}+\epsilon_{1ijkl}\lambda^{'ij}_{+}\lambda^{'kl}_{+}),\nonumber \\
\lambda^{5}_{+}=\frac{1}{\lambda_{++}}\left(\lambda^{i}_{+}\lambda_{+i5}-\lambda'_{+5}\lambda'_{++}+2\lambda'_{+i}\lambda^{'i5}_{+}+\frac{1}{2}\epsilon_{5ijkl}
\lambda^{'ij}_{+}\lambda^{'kl}_{+}+\frac{1}{8}\epsilon_{5ijkl}\lambda_{+ij}\lambda_{+kl}\right),\nonumber \\
\lambda^{'51}_{+}=\frac{1}{4\lambda_{++}}(-4\lambda^{'i1}_{+}\lambda_{+i5}-2\lambda'_{+5}\lambda^{1}_{+}+\epsilon_{15ijk}\lambda_{+ij}\lambda'_{+k}),\nonumber \\
\lambda_{+51}=\frac{1}{\lambda'_{++}}(2\lambda_{+i1}\lambda^{'i5}_{+}-\lambda^{5}_{+}\lambda'_{+1}+\epsilon_{15ijk}\lambda^{'ij}\lambda^{k}_{+}),\nonumber \\
\lambda'_{++}=\frac{1}{\lambda_{++}}(4\lambda_{+i1}\lambda^{'i1}_{+}-\lambda_{+ij}\lambda^{'ij}_{+}+\lambda^{i}_{+}\lambda'_{+i}-2\lambda^{1}_{+}\lambda'_{+1}).
\end{eqnarray}

Now from (\ref{Gauss}) we deal with the following constraints:
\begin{eqnarray}
\label{constraints}
\overline{\lambda}_{-}\Gamma^{\mu}\lambda_{-}=0, \qquad
\overline{\lambda}_{-}\Gamma^{a}\Gamma_{11}\lambda_{-}=0, \qquad
\overline{\lambda}_{-}\Gamma^{ab}\lambda_{-}=0, \qquad
\overline{\lambda}_{-}\Gamma^{\mu abcd}\lambda_{-}=0,
\end{eqnarray}
where $\mu=0,1,2/a,b,c,d=3,..,9$ with $\lambda_{-}=P_{-}\lambda$ ($P_{-}$ is given in (\ref{P_})). Thus, we can write
\begin{equation}
P_{-}=\frac{1}{2}[1+i(a^{5}-a_{5})(a_{1}a^{1}-a^{1}a_{1})].
\end{equation}

We have that $P^{T}_{-}C=CP_{-}$ where $C$ is the charge conjugation matrix. So for example the constraints $\overline{\lambda}_{-}\Gamma^{\mu}\lambda_{-}=0$ can be rewritten as follows:

\begin{eqnarray}
<\lambda\mid CP_{-}a^{1}P_{-}\mid \lambda>=0,\nonumber \\
<\lambda\mid CP_{-}a_{1}P_{-}\mid \lambda>=0,\nonumber \\
<\lambda\mid CP_{-}a^{5}P_{-}\mid \lambda>=<\lambda\mid CP_{-}a_{5}P_{-}\mid \lambda>.
\end{eqnarray}

The solution of the constraints in (\ref{constraints}) is the trivial one. This can be shown as follows.
Let us consider the following $16$ constraints:

\begin{eqnarray}\overline{\lambda}_{-}\Gamma^{13579}\lambda_{-}=0, \quad
\overline{\lambda}_{-}{\Gamma}^{13589}\lambda_{-}=0, \quad
\overline{\lambda}_{-}\Gamma^{13679}\lambda_{-}=0, \quad
\overline{\lambda}_{-}\Gamma^{13689}\lambda_{-}=0\nonumber\\
\overline{\lambda}_{-}\Gamma^{14579}\lambda_{-}=0,\quad
\overline{\lambda}_{-}\Gamma^{14679}\lambda_{-}=0,\quad
\overline{\lambda}_{-}\Gamma^{14589}\lambda_{-}=0,
\overline{\lambda}_{-}\Gamma^{14689}\lambda_{-}=0\nonumber \\
\overline{\lambda}_{-}\Gamma^{23579}\lambda_{-}=0,\quad
\overline{\lambda}_{-}\Gamma^{23589}\lambda_{-}=0,\quad
\overline{\lambda}_{-}\Gamma^{23679}\lambda_{-}=0,
\overline{\lambda}_{-}\Gamma^{23689}\lambda_{-}=0,\nonumber \\
\overline{\lambda}_{-}\Gamma^{24579}\lambda_{-}=0,\quad
\overline{\lambda}_{-}\Gamma^{24589}\lambda_{-}=0,\quad
\overline{\lambda}_{-}\Gamma^{24679}\lambda_{-}=0,
\overline{\lambda}_{-}\Gamma^{24689}\lambda_{-}=0.
\end{eqnarray}
which can be rewritten as:
\begin{eqnarray}
\label{220}
<\lambda\mid Ca^{1}((a^{5}+a_{5})-i(a_{5}a^{5}-a^{5}a_{5}))a^{2}a^{3}a^{4}\mid \lambda>=\nonumber \\
-(\lambda_{+})^{2}+(\lambda'_{5})^{2}+2i(\lambda_{+}
\lambda'_{5})=0\end{eqnarray}
\begin{eqnarray}<\lambda\mid Ca^{1}((a^{5}+a_{5})-i(a_{5}a^{5}-a^{5}a_{5}))a^{2}a^{3}a_{4}\mid \lambda>=\nonumber \\
-(\lambda_{45})^{2}+(\lambda'_{4})^{2}-2i(\lambda'_{4}\lambda_{45})=0\end{eqnarray}
\begin{eqnarray}<\lambda\mid Ca^{1}((a^{5}+a_{5})-i(a_{5}a^{5}-a^{5}a_{5}))a^{2}a_{3}a^{4}\mid \lambda>=\nonumber \\
-(\lambda_{35})^{2}+(\lambda'_{3})^{2}-2i(\lambda'_{3}\lambda_{35})=0\end{eqnarray}
\begin{eqnarray}<\lambda\mid Ca^{1}((a^{5}+a_{5})-i(a_{5}a^{5}-a^{5}a_{5}))a_{2}a^{3}a^{4}\mid \lambda>=\nonumber \\
-(\lambda_{52})^{2}+(\lambda'_{2})^{2}-2i(\lambda'_{2}\lambda_{52})=0\end{eqnarray}
\begin{eqnarray}<\lambda\mid Ca^{1}((a^{5}+a_{5})-i(a_{5}a^{5}-a^{5}a_{5}))a^{2}a_{3}a_{4}\mid \lambda>=\nonumber \\
-(\lambda_{34})^{2}+4(\lambda^{'12})^{2}-4i(\lambda_{34}\lambda^{'12})=0\end{eqnarray}
\begin{eqnarray}<\lambda\mid Ca^{1}((a^{5}+a_{5})-i(a_{5}a^{5}-a^{5}a_{5}))a_{2}a^{3}a_{4}\mid \lambda>=\nonumber \\
-(\lambda_{42})^{2}+4(\lambda^{'13})^{2}-4i(\lambda_{42}\lambda^{'13})=0\end{eqnarray}
\begin{eqnarray}<\lambda\mid Ca^{1}((a^{5}+a_{5})-i(a_{5}a^{5}-a^{5}a_{5}))a_{2}a_{3}a^{4}\mid \lambda>=\nonumber \\
-(\lambda_{23})^{2}+4(\lambda^{'14})^{2}-4i(\lambda_{23}\lambda^{'14})=0\end{eqnarray}
\begin{eqnarray}<\lambda\mid Ca^{1}((a^{5}+a_{5})-i(a_{5}a^{5}-a^{5}a_{5}))a_{2}a_{3}a_{4}\mid \lambda>=\nonumber \\
-(\lambda^{1})^{2}+4(\lambda^{'15})^{2}-4i(\lambda^{1}\lambda^{'15})=0\end{eqnarray}
\begin{eqnarray}<\lambda\mid Ca_{1}((a^{5}+a_{5})-i(a_{5}a^{5}-a^{5}a_{5}))a^{2}a^{3}a^{4}\mid \lambda>=\nonumber \\
-(\lambda_{15})^{2}+(\lambda'_{1})^{2}+2i(\lambda'_{1}\lambda_{15})=0\end{eqnarray}
\begin{eqnarray}<\lambda\mid Ca_{1}((a^{5}+a_{5})-i(a_{5}a^{5}-a^{5}a_{5}))a^{2}a^{3}a_{4}\mid \lambda>=\nonumber \\
-(\lambda_{14})^{2}+4(\lambda^{'23})^{2}+4i(\lambda_{14}\lambda^{'23})=0\end{eqnarray}
\begin{eqnarray}<\lambda\mid Ca_{1}((a^{5}+a_{5})-i(a_{5}a^{5}-a^{5}a_{5}))a^{2}a_{3}a^{4}\mid \lambda>=\nonumber \\
-(\lambda_{13})^{2}+4(\lambda^{'24})^{2}-4i(\lambda_{13}\lambda^{'24})=0\end{eqnarray}
\begin{eqnarray}<\lambda\mid Ca_{1}((a^{5}+a_{5})-i(a_{5}a^{5}-a^{5}a_{5}))a_{2}a^{3}a^{4}\mid \lambda>=\nonumber \\
-(\lambda_{12})^{2}+4(\lambda^{'34})^{2}+4i(\lambda_{12}\lambda^{'34})=0\end{eqnarray}
\begin{eqnarray}<\lambda\mid Ca_{1}((a^{5}+a_{5})-i(a_{5}a^{5}-a^{5}a_{5}))a^{2}a_{3}a_{4}\mid \lambda>=\nonumber \\
-(\lambda^{2})^{2}+4(\lambda^{'52})^{2}-4i(\lambda^{2}\lambda^{'52})=0\end{eqnarray}
\begin{eqnarray}<\lambda\mid Ca_{1}((a^{5}+a_{5})-i(a_{5}a^{5}-a^{5}a_{5}))a_{2}a^{3}a_{4}\mid \lambda>=\nonumber \\
-(\lambda^{3})^{2}+4(\lambda^{'53})^{2}+4i(\lambda^{3}\lambda^{'53})=0\end{eqnarray}
\begin{eqnarray}<\lambda\mid Ca_{1}((a^{5}+a_{5})-i(a_{5}a^{5}-a^{5}a_{5}))a_{2}a_{3}a^{4}\mid \lambda>=\nonumber \\
-(\lambda^{4})^{2}+4(\lambda^{'45})^{2}+4i(\lambda^{4}\lambda^{'45})=0\end{eqnarray}
\begin{eqnarray}
\label{235}
<\lambda\mid Ca_{1}((a^{5}+a_{5})-i(a_{5}a^{5}-a^{5}a_{5}))a_{2}a_{3}a_{4}\mid \lambda>=\nonumber \\
-(\lambda^{5})^{2}+(\lambda'_{+})^{2}-2i(\lambda^{5}\lambda'_{+})=0
\end{eqnarray}
and the expression for $\mid \lambda>$ is given in (\ref{spinor}). According to\textit{ Comment 2} in \cite{GPN}--\cite{GPN1} we choose all the $\lambda$ s and
$\lambda'$ s to be real. However the Dirac spinor becomes complex in a general Lorentz frame. Then the relations (\ref{220})-(\ref{235}) imply that all the components of  $\mid \lambda>$ are zero and the solution of the constraints in (2.16) is the trivial one. It is interesting to note that $\lambda_{-}$ corresponds to $\theta_{-}$ which is a gauge degree of freedom as pointed out in \cite{GPR}.

Thus, the novel nontrivial solution (\ref{soln}) found for one of the spinor fields actually proves that the pure spinor formalism can be successfully applied to non-relativistic systems with kappa symmetry as is the case of the IIA D2-brane.

%
%
\section{Conclusions}

This paper is the first attempt to extend the application of the pure
spinor formalism to non-relativistic systems with kappa symmetry expecting to
draw some conclusions for the general case as well. This can very useful especially in cases
where it is difficult to solve the pure spinor constraints in the general case.
We treated the non-relativistic IIA D2 brane in the framework of this formalism \cite {JH} and we derived the
fermionic constraints corresponding to the rescaled fermionic
coordinates. We introduced two commuting spinor fields each one
corresponding to a fermionic coordinate. The nilpotency of the
BRST charge leads to a set of constraints for the two spinor fields
including pure spinor constraints. Nontrivial solutions are found
for the spinor field $\lambda_{+}$ which corresponds to the
fermionic coordinate $\theta_{+}$. It is interesting to note that this
solution can be written as the sum of two pure spinors $\lambda_{1+}=P_+\lambda_+$ and
$\lambda_{2+}=P_-\lambda_+$ where $P_\pm= \frac{1}{2}(1\pm\Gamma_0\Gamma_1\Gamma_2)$.
The solution for the spinor field $\lambda_{-}$ corresponding to $\theta_{-}$
which, according to the proof given in \cite{GPR} constitutes a gauge
degree of freedom, is the trivial one. 
So for example the expression of the BRST charge
 for $\theta_{-}$= 0 in the Schr\"odinger representation is given by
 $Q=\int d^2\sigma[-i(\lambda_{1+}+\lambda_{2+})^\alpha\frac{\partial}{\partial \theta_{+}^\alpha} + 2T_{NR}\epsilon_{ijk}(\bar{\theta}_{+}\Gamma^{i}\lambda_{1+})\partial_{1}x^{j}\partial_{2}x^{k}]$.
 Cohomological issues concerning the BRST charge are currently under investigation \cite {BF},\cite {MAX},\cite {Xu}, \cite {AXU}.
 This study can also be performed for more general manifolds and for general dimensions as well. We would
 like to finally mention that the treatment of the relativistic Dp-brane in the framework of the pure
spinor formalism has been reported in \cite {AG}.


\section*{Acknowledgments}

We gratefully acknowledge support from ``CONACYT grant A1-S-38041, {\it Aspectos gravitatorios de la correspondencia hologr\'afica}Ó . AHA acknowledges a VIEP-BUAP research grant, he thanks as well SNI and PRODEP for partial financial support, while JEP is grateful to the Instituto de F\'\i sica y Matem\'aticas, at UMSNH, in Morelia, for the warm hospitality and the stimulating atmosphere generated during his visit at the Institute.


\section{Appendix A}

For the matrices $\Gamma^{m}\, (m=0,..,9)$ we use the Majorana representation ($\Gamma^{0}\,$ is real antisymmetric, $\Gamma^{i}\, (i=1,..,9)$ are real symmetric)
\begin{equation}\{\Gamma^{m},\Gamma^{n}\}=2\eta^{mn}
\end{equation}
where
$\eta^{mn}=diag(-1,1,..,1)$ and $m,n=0,..,9;$ and
\begin{eqnarray} C=\Gamma_{0},\nonumber \\
(\Gamma^{m})^{T}=-\Gamma_{0}\Gamma^{m}\Gamma^{-1}_{0}\,.
\end{eqnarray}

A basis for the $32\times 32$ matrices is given by (see Appendix B in \cite{HEN} for instance):
\begin{equation} B=\left\{I,\Gamma^{A}\Gamma^{AB},\Gamma^{ABC},\Gamma^{ABCD},\Gamma^{ABCDE},\Gamma^{ABCD}\Gamma_{11},\Gamma^{AB}\Gamma_{11},\Gamma^{A}\Gamma_{11},\Gamma_{11}\right\},
\end{equation}
with $A<B<C<D<E\,.$

For the differential form $d\theta$ we use the convention given in
\cite{APS}:
\begin{eqnarray}
d\theta=d\sigma^{\mu}\partial_{\mu}\theta=-\partial_{\mu}\theta d\sigma^{\mu}\,.
\end{eqnarray}

The following identities hold for $\lambda_{-}=P_{-}\lambda$:
\begin{eqnarray}\Gamma_{\mu\nu}\lambda_{-}=-\epsilon_{\mu\nu\rho}\Gamma^{\rho}\lambda_{-}\nonumber \\
\Gamma_{\mu\nu\rho}\lambda_{-}=-\epsilon_{\mu\nu\rho}\lambda_{-}\end{eqnarray}
$\epsilon_{012}=1.$

\section{Appendix B}

\begin{eqnarray}
F_{\mu}=2[p_{\mu}-E^{i}(\overline{\theta}_{-}\Gamma_{11}\Gamma_{\mu}\partial_{i}\theta_{+}+\overline{\theta}_{+}
\Gamma_{11}\Gamma_{\mu}\partial_{i}\theta_{-})]+2T_{NR}\epsilon_{0jk}[-2(\overline{\theta}_{+}\Gamma_{\mu\nu}\partial_{j}\theta_{+})R^{\nu}_{k}+
\nonumber \\
(\overline{\theta}_{+}\Gamma_{a\mu}\partial_{j}\theta_{-}+\overline{\theta}_{-}\Gamma_{a\mu}\partial_{j}\theta_{+})u^{a}_{k}]+\frac{T_{NR}}{96}
\epsilon_{0jk}[(\overline{\theta}_{-}\Gamma^{\nu}\partial_{k}\theta_{-})(-89(\overline{\theta}_{+}\Gamma_{\mu\nu}\partial_{j}\theta_{+})
+\nonumber \\
22(\overline{\theta}_{+}\partial_{j}\theta_{+})\eta_{\mu\nu})+(\overline{\theta}_{+}\Gamma^{\nu}\partial_{k}\theta_{+})(83(\overline{\theta}_{-}
\Gamma_{\mu\nu}\partial_{j}\theta_{-})+6(\overline{\theta}_{-}\partial_{j}\theta_{-})\eta_{\mu\nu})-2(\overline{\theta}_{-}\Gamma^{a}\partial
_{k}\theta_{+}+\nonumber \\
\overline{\theta}_{+}\Gamma^{a}\partial_{k}\theta_{-})(47(\overline{\theta}_{+}\Gamma_{\mu
a }\partial_{j}\theta_{-})+23(\overline{\theta}_{-}\Gamma_{\mu a
}\partial_{j}\theta_{+}))-5(\overline{\theta}_{+}\Gamma_{\mu}\Gamma_{\nu
a
}\partial_{j}\theta_{-})(\partial_{k}\overline{\theta}_{+}\Gamma^{\nu
a }\theta_{-})-\nonumber \\
\frac{1}{3}(\partial_{k}\overline{\theta}_{+}\Gamma^{a}\theta_{-})(37(\partial_{j}\overline{\theta}_{-}\Gamma_{\mu
a }\theta_{+})-66(\partial_{j}\overline{\theta}_{+}\Gamma_{\mu a
}\theta_{-}))-(\partial_{j}\theta_{-}\Gamma^{a}\theta_{+})\nonumber
\\
 (21(\partial_{k}\overline{\theta}_{+}\Gamma_{\mu a
}\theta_{-})-2(\partial_{k}\theta_{-}\Gamma_{\mu
a}\theta_{+}))-2(\overline{\theta}_{+}\Gamma_{11}\partial_{ê}\theta_{-}+\overline{\theta}_{-}\Gamma_{11}\partial_{ê}\theta_{+})\nonumber
\\
(25(\overline{\theta}_{+}\Gamma_{11}\Gamma_{\mu}\partial_{j}\theta_{-})+31(\overline{\theta}_{-}\Gamma_{11}\Gamma_{\mu}\partial_{j}\theta_{+}))
+2(\overline{\theta}_{-}\Gamma_{11}\Gamma^{\nu}\partial_{k}\theta_{+}+\overline{\theta}_{+}\Gamma_{11}\Gamma^{\nu}\partial_{k}\theta_{-})\nonumber
\\
(\overline{\theta}_{+}\Gamma_{11}\Gamma_{\mu}\Gamma_{\nu}\partial_{j}\theta_{-}+2(\overline{\theta}_{-}\Gamma_{11}\partial_{j}\theta_{+})\eta_{\mu\nu})-
(\partial_{j}\overline{\theta}_{-}\Gamma_{11}\Gamma^{\nu}\theta_{-})(64(\partial_{k}\overline{\theta}_{+}\Gamma_{11}\theta_{-})-\nonumber
\\
11(\partial_{k}\overline{\theta}_{+}\Gamma_{11}\Gamma_{\mu\nu}\theta_{-})+46(\partial_{k}\overline{\theta}_{-}\Gamma_{11}\theta_{+})\eta_{\mu\nu}-
2(\partial_{k}\overline{\theta}_{-}\Gamma_{11}\Gamma_{\nu}\Gamma_{\mu}\theta_{+}))+(\overline{\theta}_{-}\Gamma_{11}\Gamma_{\nu}\partial_{k}\theta_{+})
\nonumber \\
(9(\overline{\theta}_{-}\Gamma_{11}\Gamma_{\nu}\Gamma_{\mu}\theta_{+})+34(\overline{\theta}_{-}\Gamma_{11}\partial_{j}\theta_{+})\eta_{\mu\nu}
+23(\overline{\theta}_{+}\Gamma_{11}\partial_{j}\theta_{-})\eta_{\mu\nu})-7(\partial_{j}\overline{\theta}_{-}\Gamma_{11}\Gamma_{a}\theta_{-})\nonumber
\\
(\partial_{k}\overline{\theta}_{+}\Gamma_{11}\Gamma_{\mu
a}\theta_{+})+5(\partial_{k}\overline{\theta}_{+}\Gamma_{11}\Gamma^{\nu
a }\Gamma_{\mu}\theta_{+})(\overline{\theta}_{-}\Gamma_{\nu
a}\Gamma_{11}\partial_{j}\theta_{-})+(\overline{\theta}_{-}\Gamma^{\nu
a }\Gamma_{11}\Gamma_{\mu}\partial_{j}\theta_{-})\nonumber \\
(\partial_{k}\overline{\theta}_{+}\Gamma_{\nu
a}\Gamma_{11}\theta_{+})-(\partial_{k}\overline{\theta}_{+}\Gamma_{11}\Gamma^{a}\theta_{+})(\partial_{j}\overline{\theta}_{-}\Gamma_{11}\Gamma_{\mu
a }\theta_{-})+(\overline{\theta}_{+}\Gamma_{\nu
a}\partial_{j}\theta_{-})(\overline{\theta}_{-}\Gamma^{\nu
a}\Gamma_{\mu}\partial_{k}\theta_{+})]
\end{eqnarray}

\begin{eqnarray}
F_{a}=-2E_{i}u_{a}^{i}+2T_{NR}\epsilon_{0jk}(\overline{\theta}_{+}\Gamma_{11}\partial_{j}\theta_{-}+\overline{\theta}_{-}\Gamma_{11}
\partial_{j}\theta_{+})u_{ak}+\frac{T_{NR}}{3}\epsilon_{0jk}[-\frac{1}{16}(\overline{\theta}_{+}\Gamma^{b}\partial_{j}\theta_{-})\nonumber \\
(10(\overline{\theta}_{-}\Gamma_{ab}\Gamma_{11}\partial_{k}\theta_{+})-(\overline{\theta}_{+}\Gamma_{ab}\Gamma_{11}\partial_{k}\theta_{-})+3(
\partial_{k}\theta_{-}\Gamma_{11}\theta_{+})\eta_{ab}+45(\partial_{k}\overline{\theta}_{+}\Gamma_{11}\theta_{-})\eta_{ab})+\nonumber \\
\frac{3}{16}(\overline{\theta}_{-}\Gamma^{b}\partial_{j}\theta_{+})(-8(\partial_{k}\overline{\theta}_{+}\Gamma_{ab}\Gamma_{11}\theta_{-})+12
(\overline{\theta}_{-}\Gamma_{11}\partial_{k}\theta_{+})\eta_{ab}-(\overline{\theta}_{+}\Gamma_{11}\partial_{k}\theta_{-})\eta_{ab})+\nonumber \\
\frac{7}{8}(\overline{\theta}_{+}\Gamma^{\mu}\partial_{k}\theta_{+})(\partial_{j}\overline{\theta}_{-}\Gamma_{\mu a}\Gamma_{11}\theta_{-})+\frac{1}{48}(\overline{\theta}_{+}\Gamma_{11}\Gamma^{\mu}\partial_{k}\theta_{-})(5(\partial_{j}\overline{\theta}_{+}\Gamma_{\mu a}\theta_{-})-6(\partial_{j}\overline{\theta}_{-}\Gamma_{\mu a}\theta_{+}))-\nonumber \\
\frac{1}{16}(\partial_{j}\overline{\theta}_{-}\Gamma_{11}\Gamma^{b}\theta_{-})(13(\partial_{k}\overline{\theta}_{+}\Gamma_{ab}\theta_{+})+3(\overline{\theta
}_{+}\partial_{k}\theta_{+})\eta_{ab})-\frac{1}{32}(\partial_{j}\overline{\theta}_{-}\Gamma_{abc}\theta_{+})\nonumber \\
(2(\partial_{k}\overline{\theta}_{+}\Gamma^{bc}
\Gamma_{11}\theta_{-})+\partial_{k}\overline{\theta}_{-}\Gamma^{bc}\Gamma_{11}\theta_{+})+\frac{1}{16}(\partial_{j}\overline{\theta}_{-}\Gamma^{\mu b}\theta_{+})(-2(\partial_{k}\overline{\theta}_{-}\Gamma_{\mu ab}\Gamma_{11}\theta_{+})\nonumber \\
-6(\partial_{k}\overline{\theta}_{+}\Gamma_{11}\Gamma_{\mu ab}\theta_{-})+9(\overline{\theta}_{-}\Gamma_{11}\Gamma_{\mu}\partial_{k}\theta_{+})\eta_{ab}+(\overline{\theta}_{+}\Gamma_{11}\Gamma_{\mu}\partial_{k}
\theta_{-})\eta_{ab})-\frac{9}{16}(\overline{\theta}_{-}\Gamma^{\mu}\partial_{j}\theta_{-})\nonumber \\
(\overline{\theta}_{+}\Gamma_{\mu a}\Gamma_{11}\partial_{k}\theta_{+})+\frac{3}{8}(\overline{\theta}_{+}\Gamma^{\mu b}\Gamma_{11}\partial_{k}\theta_{+})(\partial_{j}\overline{\theta}_{-}\Gamma_{\mu ab}\theta_{-})+\frac{117}{64}(\overline{\theta}_{-}\partial_{j}\theta_{-})(\overline{\theta}_{+}\Gamma_{11}\Gamma_{a}\partial_{k}\theta_{+})+\nonumber \\
\frac{137}{64}(\overline{\theta}_{-}\Gamma_{ab}\partial_{j}\theta_{-})(\overline{\theta}_{+}\Gamma_{11}\Gamma^{b}\partial_{k}\theta_{+})+
\frac{1}{16}(\overline{\theta}_{-}\Gamma^{\mu b}\partial_{j}\theta_{+})(\partial_{k}\overline{\theta}_{-}\Gamma_{11}\Gamma_{\mu ab}\theta_{+})-
\frac{1}{16}(\overline{\theta}_{-}\Gamma_{11}\Gamma^{\mu b}\partial_{j}\theta_{-})\nonumber \\
(\partial_{k}\overline{\theta}_{+}\Gamma_{\mu ab}\theta_{+})-\frac{1}{16}(\overline{\theta}_{-}\Gamma^{bc}\partial_{k}\theta_{-})(\overline{\theta}_{+}\Gamma_{abc}\Gamma_{11}\partial_{j}\theta_{+})]
\end{eqnarray}

\begin{eqnarray}
F_{ab}=T_{NR}\epsilon_{0jk}u_{aj}u_{bk}+\frac{T_{NR}}{16}\epsilon_{0jk}[(\overline{\theta}_{+}\Gamma^{d}\partial_{j}\theta_{-})(-\frac{14}{3}
(\partial_{k}\overline{\theta}_{-}\Gamma_{ab}\Gamma_{d}\theta_{+})+\nonumber \\
\frac{14}{3}(\partial_{k}\overline{\theta}_{-}\Gamma_{d}\Gamma_{ab}\theta_{+})+\frac{35}{6}(\partial_{k}\overline{\theta}_{+}\Gamma_{ab}\Gamma_{d}\theta_{-})-\frac{15}{2}(
\partial_{k}\overline{\theta}_{+}\Gamma_{d}\Gamma_{ab}\theta_{-})+\frac{16}{3}(\overline{\theta}_{+}\Gamma_{b}\partial_{k}\theta_{-})\eta_{ad}+\nonumber \\
\frac{32}{3}(\overline{\theta}_{-}\Gamma_{b}\partial_{k}\theta_{+})\eta_{ad})+\frac{1}{6}(\overline{\theta}_{-}\Gamma^{d}\partial_{j}\theta_{+})(7(\partial_{k}
\overline{\theta}_{+}\Gamma_{d}\Gamma_{ab}\theta_{-})+35(\partial_{k}\overline{\theta}_{+}\Gamma_{ab}\Gamma_{d}\theta_{-})+\nonumber \\
4(\partial_{k}\overline{\theta}_{-}\Gamma_{d}\Gamma_{ab}\theta_{+})-33(\partial_{k}\overline{\theta}_{-}\Gamma_{ab}\Gamma_{d}\theta_{+})-32(\overline{\theta}
_{-}\Gamma_{a}\partial_{k}\theta_{+})\eta_{bd})+\frac{1}{6}(\overline{\theta}_{-}\Gamma_{11}\Gamma^{\mu}\partial_{k}\theta_{+})\nonumber \\
(-4(\partial_{j}\overline{\theta}_{+}\Gamma_{11}\Gamma_{\mu ab}\theta_{-})+37(\partial_{j}\overline{\theta}_{-}\Gamma_{11}\Gamma_{\mu ab}\theta_{-}))+\frac{15}{4}(\overline{\theta}_{+}\Gamma^{\mu d}\Gamma_{11}\partial_{k}\theta_{+})(\partial_{j}\overline{\theta}_{-}\Gamma_{11}\Gamma_{\mu d}\Gamma_{ab}\theta_{-})+\nonumber \\
\frac{25}{12}(\partial_{j}\overline{\theta}_{-}\Gamma^{\mu d}\theta_{+})(\partial_{k}\overline{\theta}_{+}\Gamma_{\mu d}\Gamma_{ab}\theta_{-})-
\frac{77}{12}(\overline{\theta}_{+}\Gamma^{d}\Gamma_{11}\partial_{k}\theta_{+})(\partial_{j}\overline{\theta}_{-}\Gamma_{11}\Gamma_{d}\Gamma_{ab}\theta_{-})-\nonumber \\
\frac{35}{4}(\partial_{j}\overline{\theta}_{-}\Gamma_{11}\theta_{+})(\partial_{k}\overline{\theta}_{+}\Gamma_{11}\Gamma_{ab}\theta_{-})-\frac{45}{4}(\overline{\theta}
_{-}\Gamma_{ab}\partial_{j}\theta_{-})(\overline{\theta}_{+}\partial_{k}\theta_{+})+\frac{37}{4}(\overline{\theta}_{+}\Gamma^{\mu}\partial_{k}\theta_{+})\nonumber \\
(\partial_{j}\overline{\theta}_{-}\Gamma_{\mu ab}\theta_{-})+\frac{1}{6}(\overline{\theta}_{-}\Gamma_{11}\Gamma^{d}\partial_{j}\theta_{-})(5(\partial_{k}\overline{\theta}_{+}\Gamma_{11}\Gamma_{ab}\Gamma_{d}
\theta_{+})+39(\partial_{k}\overline{\theta}_{+}\Gamma_{d}\Gamma_{ab}\Gamma_{11}\theta_{+}))-\nonumber \\
3(\overline{\theta}_{-}\Gamma_{11}\partial_{k}\theta_{+})(\partial_{j}\overline{\theta}_{-}\Gamma_{11}\Gamma_{ab}\theta_{+})+(\overline{\theta}_{-}\Gamma^{\mu d}\partial_{j}\theta_{+})(\partial_{k}\overline{\theta}_{-}\Gamma_{ab}\Gamma_{\mu d}\theta_{+})-(\overline{\theta}_{-}\Gamma_{df}\partial_{k}\theta_{-})\nonumber \\
(\overline{\theta}_{+}\Gamma^{d}\Gamma_{ab}\Gamma^{f}\partial_{j}\theta_{+})-\frac{1}{4}(\overline{\theta}_{+}\Gamma_{11}\Gamma^{\mu}\partial_{k}\theta_{-})(
\partial_{j}\overline{\theta}_{+}\Gamma_{11}\Gamma_{\mu ab}\theta_{-})-\frac{1}{2}(\overline{\theta}_{-}\partial_{j}\theta_{-})(\partial_{k}\overline{\theta}_{+}\Gamma_{ab}\theta_{+})-\nonumber \\
\frac{1}{6}(\overline{\theta}_{-}\Gamma_{11}\Gamma^{\mu d}\partial_{j}\theta_{-})(\partial_{k}\overline{\theta}_{+}\Gamma_{11}\Gamma_{\mu ab}\Gamma_{d}\theta_{+})+\frac{1}{6}(\partial_{k}\overline{\theta}_{+}\Gamma_{df}\Gamma_{11}\theta_{-})(\partial_{j}\overline{\theta}_{-}\Gamma_{11}\Gamma^{d}
\Gamma_{ab}\Gamma^{f}\theta_{+})-\nonumber \\
4(\overline{\theta}_{-}\Gamma^{\mu}\partial_{j}\theta_{-})(\partial_{k}\overline{\theta}_{+}\Gamma_{\mu ab}\theta_{+})]\end{eqnarray}

\begin{eqnarray}
\overline{F}=2T_{NR}\epsilon_{0jk}\Big[\widetilde{{\cal F}}_{jk}^{(1)}-\frac{1}{3}((\overline{\theta}_{-}\Gamma^{\mu}\partial_{k}\theta_{-})(\overline{\theta}_{+}\Gamma_{\mu}\Gamma_{11}\partial_{j}\theta_{-}+
\overline{\theta}_{-}\Gamma_{\mu}\Gamma_{11}\partial_{j}\theta_{+})+\nonumber  \\
(\overline{\theta}_{-}\Gamma^{a}\Gamma_{11}\partial_{j}\theta_{-})(\overline{\theta}_{-}\Gamma_{a}\partial_{k}\theta_{+}+\overline{\theta}_{+}\Gamma_{a}
\partial_{k}\theta_{-}))\Big]\end{eqnarray}

\begin{eqnarray}
\overline{F}_{\nu}=-4E^{i}R_{\nu i}+4T_{NR}\epsilon_{0jk}R_{\nu k}(\overline{\theta}_{+}\Gamma_{11}\partial_{j}\theta_{-}+\overline{\theta}
{-}\Gamma_{11}\partial_{j}\theta_{+})+\frac{1}{48}T_{NR}\epsilon_{0jk}[-(\overline{\theta}_{-}\Gamma^{\mu
a }\partial_{j}\theta_{+})\nonumber \\
(16(\partial_{k}\overline{\theta}_{-}\Gamma_{11}\Gamma_{a}\theta_{-})\eta_{\mu\nu}-5(\partial_{k}\overline{\theta}_{-}\Gamma_{\mu\nu
a
}\Gamma_{11}\theta_{-}))-12(\partial_{k}\overline{\theta}_{-}\Gamma_{11}\Gamma^{a}\theta_{-})(\overline{\theta}_{+}\Gamma_{\nu
a }\partial_{j}\theta_{-})-\nonumber \\
(\overline{\theta}_{-}\Gamma^{\mu}\partial_{j}\theta_{-})(-37(\partial_{k}\overline{\theta}_{+}\Gamma_{\mu\nu}\Gamma_{11}\theta_{-})+
41(\partial_{k}\overline{\theta}_{+}\Gamma_{11}\theta_{-})\eta_{\mu\nu}+18(\partial_{k}\overline{\theta}_{-}\Gamma_{11}\theta_{+})\eta_{\mu\nu})+
\nonumber \\
(\partial_{j}\overline{\theta}_{-}\Gamma_{\nu
a}\Gamma_{b}\theta_{-})(-5(\partial_{k}\overline{\theta}_{+}\Gamma^{b}\Gamma^{a}\Gamma_{11}\theta_{-})+2(\partial_{k}\overline{\theta}_{-}
\Gamma^{b}\Gamma^{a}\Gamma_{11}\theta_{+}))-(\overline{\theta}_{-}\Gamma^{ab}\partial_{k}\theta_{-})\nonumber
\\
(7(\overline{\theta}_{-}\Gamma_{\nu
ab}\Gamma_{11}\partial_{j}\theta_{+})+2(\overline{\theta}_{+}\Gamma_{\nu
ab
}\Gamma_{11}\partial_{j}\theta_{-}))+(\partial_{j}\overline{\theta}_{+}\Gamma_{11}\Gamma^{\mu}\theta_{-})(70(\overline{\theta}_{-}\partial
_{k}\theta_{-})\eta_{\mu\nu}+\nonumber \\
59(\partial_{k}\overline{\theta}_{-}\Gamma_{\mu\nu}\theta_{-}))+(\partial_{j}\overline{\theta}_{-}\Gamma_{\nu
a}\Gamma_{11}\theta_{-})(49(\overline{\theta}_{-}\Gamma^{a}\partial_{k}\theta_{+})+24(\overline{\theta}_{+}\Gamma^{a}\partial_{k}\theta_{-}))-
\nonumber \\
5(\partial_{k}\overline{\theta}_{+}\Gamma^{\mu
a}\Gamma_{\nu}\theta_{-})(\overline{\theta}_{-}\Gamma_{11}\Gamma_{\mu
a }\partial_{j}\theta_{-})]\end{eqnarray}

\begin{eqnarray}
\overline{F}_{a}=4(p_{a}-E^{i}(\overline{\theta}_{-}\Gamma_{11}\Gamma_{a}\partial_{i}\theta_{-}))-2T_{NR}\epsilon_{0jk}[(\overline{\theta}
_{-}\Gamma_{\mu a}\partial_{k}\theta_{+}+\overline{\theta}_{+}\Gamma_{\mu a}\partial_{k}\theta_{-})R^{\mu}_{j}\nonumber \\
-(\overline{\theta}_{-}\Gamma_{ab}\partial_{k}\theta_{-})u^{b}_{j}]+\frac{2}{3}T_{NR}\epsilon_{0jk}[\frac{1}{16}(\overline{\theta}_{-}
\Gamma^{\mu}\partial_{j}\theta_{-})(25(\partial_{k}\overline{\theta}_{+}\Gamma_{\mu
a }\theta_{-})+19(\partial_{k}\theta_{-}\Gamma_{\mu a}\theta_{+}))+\nonumber \\
\frac{1}{8}(\overline{\theta}_{-}\partial_{j}\theta_{-})(21(\partial_{k}\overline{\theta}_{+}\Gamma_{a}\theta_{-})+9(\partial_{k}
\overline{\theta}_{-}\Gamma_{a}\theta_{+}))-\frac{1}{32}(\partial_{j}\overline{\theta}_{-}\Gamma_{\mu
ab }\theta_{-})(5(\overline{\theta}_{-}\Gamma^{\mu
b}\partial_{k}\theta_{+})-\nonumber \\
2(\overline{\theta}_{+}\Gamma^{\mu
b}\partial_{k}\theta_{-}))+\frac{1}{32}(\overline{\theta}_{-}\Gamma_{ab}\partial_{k}\theta_{-})(7(\overline{\theta}_{-}\Gamma^{b}\partial_{j}
\theta_{+})+12(\overline{\theta}_{+}\Gamma^{b}\partial_{j}\theta_{-}))+\frac{1}{16}(\partial_{j}\overline{\theta}_{-}\Gamma_{11}\Gamma_{\mu
a }\theta_{-})\nonumber \\
(3(\partial_{k}\overline{\theta}_{-}\Gamma_{11}\Gamma^{\mu}\theta_{+})+10(\partial_{k}\overline{\theta}_{+}\Gamma_{11}\Gamma^{\mu}\theta_{-}))
+\frac{1}{16}(\overline{\theta}_{-}\Gamma_{a}\Gamma_{11}\partial_{j}\theta_{-})(29(\overline{\theta}_{+}\Gamma_{11}\partial_{k}\theta_{-})+\nonumber
\\
20(\overline{\theta}_{-}\Gamma_{11}\partial_{k}\theta_{+}
))+\frac{1}{32}(\overline{\theta}_{-}\Gamma_{11}\Gamma^{b}\partial_{k}\theta_{-})(2(\overline{\theta}_{+}\Gamma_{11}\Gamma_{ab}\partial_{j}\theta
_{-})-\overline{\theta}_{-}\Gamma_{11}\Gamma^{ab}\partial_{j}\theta_{+})+\nonumber
\\
\frac{1}{32}(\partial_{j}\overline{\theta}_{-}\Gamma_{11}\Gamma_{abc}\theta_{-})(\overline{\theta}_{+}\Gamma_{11}\Gamma^{bc}\partial_{k}\theta_{-}-
5(\overline{\theta}_{-}\Gamma_{11}\Gamma^{bc}\partial_{k}\theta_{+}))+\frac{1}{32}(\overline{\theta}_{-}\Gamma^{bc}\partial_{j}\theta_{-})\nonumber
\\
(7(\partial_{k}\overline{\theta}_{+}\Gamma_{abc}\theta_{-})+2(\partial_{k}\overline{\theta}_{-}\Gamma_{abc}\theta_{+}))-\frac{5}{32}(\partial_{k}
\overline{\theta}_{+}\Gamma_{11}\Gamma_{\mu
ab}\theta_{-})(\overline{\theta}_{-}\Gamma_{11}\Gamma^{\mu
b}\partial_{j}\theta_{-})]\end{eqnarray}

\begin{eqnarray}
\overline{F}_{\nu a}=4T_{NR}\epsilon_{0jk}R_{\nu j}u_{ak}+\frac{1}{48}T_{NR}\epsilon_{0jk}[(\overline{\theta}_{-}\Gamma^{\mu}
\partial_{j}\theta_{+})(37(\overline{\theta}_{-}\Gamma_{a}\partial_{k}\theta_{+})\eta_{\mu\nu}+\nonumber \\
34(\overline{\theta}_{+}\Gamma_{a}\partial_{k}\theta_{-})
\eta_{\mu\nu}-5(\overline{\theta}_{-}\Gamma_{\mu\nu a}\partial_{k}\theta_{+})+6(\overline{\theta}_{+}\Gamma_{\mu\nu a}\partial_{k}\theta_{-}))+
(\overline{\theta}_{-}\Gamma_{11}\Gamma^{b}\partial_{j}\theta_{-})\nonumber \\
(5(\partial_{k}\overline{\theta}_{+}\Gamma_{11}\Gamma_{\nu}(\eta_{ab}+
\Gamma_{ba})\theta_{-})-2(\partial_{k}\overline{\theta}_{-}\Gamma_{11}\Gamma_{\nu ba}\theta_{+}))-(\overline{\theta}_{-}\Gamma^{b}\partial_{k}\theta_{+})\nonumber \\
(-19(\partial_{j}\overline{\theta}_{-}\Gamma_{\nu ba}\theta_{-})+9(\partial_{j}\overline{\theta}_{-}\Gamma_{\nu}\theta_{-})\eta_{ab})+
(\overline{\theta}_{-}\Gamma^{b\mu}\Gamma_{11}\partial_{j}\theta_{-})(\partial_{k}\overline{\theta}_{+}\Gamma_{\mu}\Gamma_{11}\Gamma
_{\nu a}\Gamma_{b}\theta_{-})+\nonumber \\
5(\overline{\theta}_{-}\Gamma^{\mu}\Gamma_{11}\partial_{k}\theta_{+})(\partial_{j}\overline{\theta}_{-}\Gamma_{\mu}\Gamma_{11}\Gamma_{\nu a}\theta_{-})-(\overline{\theta}_{-}\Gamma^{bc}\Gamma_{11}\partial_{k}\theta_{+})(\partial_{j}\overline{\theta}_{-}\Gamma_{11}\Gamma_{\nu abc}\theta_{-})+\nonumber \\
(\partial_{k}\overline{\theta}_{+}\Gamma^{b\mu}\theta_{-})(\partial_{j}\overline{\theta}_{-}\Gamma_{\mu}\Gamma_{\nu a}\Gamma_{b}\theta_{-})-
5(\overline{\theta}_{-}\Gamma_{11}\partial_{k}\theta_{+})(\overline{\theta}_{-}\Gamma_{\nu a}\Gamma_{11}\partial_{j}\theta_{-})-
22(\overline{\theta}_{+}\Gamma^{b}\partial_{j}\theta_{-})\nonumber \\
(\overline{\theta}_{-}\Gamma_{\nu ba}\partial_{k}\theta_{-})+2(\overline{\theta}_{+}\Gamma_{11}\Gamma^{\mu}\partial_{k}\theta_{-})(-
\partial_{j}\overline{\theta}_{-}\Gamma_{11}\Gamma_{\mu\nu a}\theta_{-}+2(\partial_{j}\overline{\theta}_{-}\Gamma_{11}\Gamma_{a}\theta_{-})\eta_{\mu\nu})+\nonumber \\
2(\partial_{k}\overline{\theta}_{-}\Gamma^{b\mu}\theta_{+})(2(\partial_j\overline{\theta}_{-}\Gamma_{\mu\nu}\theta_{-})\eta_{ab}+(\partial_{j}
\overline{\theta}_{-}\Gamma_{\mu\nu ba}\theta_{-})-2(\partial_{j}\overline{\theta}_{-}\Gamma_{ba}\theta_{-})\eta_{\mu\nu}-\nonumber \\
(\partial_{j}\overline{\theta}_{-}\theta_{-})\eta_{\mu\nu}\eta_{ab})-2(\partial_{j}\overline{\theta}_{-}\Gamma_{11}\theta_{+})(\partial_{k}
\overline{\theta}_{-}\Gamma_{11}\Gamma_{\nu a}\theta_{-})+(\partial_{k}\overline{\theta}_{-}\Gamma^{bc}\theta_{-})\nonumber \\
(\partial_{j}\overline{\theta}
_{-}\Gamma_{\nu abc}\theta_{+}-2(\partial_{j}\overline{\theta}_{+}\Gamma_{\nu abc}\theta_{-}))]\end{eqnarray}

\begin{eqnarray}
\overline{F}^{\nu ab}=\frac{1}{96}T_{NR}\epsilon_{0jk}[(\partial_{k}\overline{\theta}_{-}\Gamma^{\mu}\Gamma_{11}\Gamma^{\nu ab}\Gamma^{d}\theta_{-})(4(\partial_{j}\overline{\theta}_{-}\Gamma_{d\mu}\theta_{+})-\partial_{j}\overline{\theta}_{+}\Gamma_{d\mu}\theta_{-})-\nonumber \\
6(\partial_{k}\overline{\theta}_{-}\Gamma^{d}\theta_{+})(\partial_{j}\overline{\theta}_{-}(\Gamma_{d}\Gamma^{\nu ab}+3\Gamma^{\nu ab}\Gamma_{d})\Gamma_{11}\theta_{-})+3(\overline{\theta}_{-}\Gamma^{d}\partial_{k}\theta_{+})\nonumber \\
(\partial_{j}\overline{\theta}_{-}(\Gamma_{d}\Gamma^{\nu ab}+6\Gamma^{\nu ab}\Gamma_{d})\Gamma_{11}\theta_{-})+(\partial_{k}\overline{\theta}
_{+}\Gamma_{dc}\Gamma_{11}\theta_{-})(\partial_{j}\overline{\theta}_{-}\Gamma^{d}\Gamma^{\nu ab}\Gamma^{c}\theta_{-})-\nonumber \\
3(\partial_{k}\overline{\theta}_{+}\Gamma_{11}\theta_{-})(\partial_{j}\overline{\theta}_{-}\Gamma^{\nu ab}\theta_{-})+
3(\partial_{j}\overline{\theta}_{-}\Gamma^{\mu}\theta_{-})(\partial_{k}\overline{\theta}_{+}\Gamma_{\mu}\Gamma_{11}\Gamma^{\nu ab}\theta_{-})-\nonumber \\
(\overline{\theta}_{-}\Gamma_{cd}\partial_{j}\theta_{-})(\partial_{k}\overline{\theta}_{+}\Gamma^{c}\Gamma_{11}\Gamma^{\nu ab}\Gamma^{d}\theta_{-})+(\overline{\theta}_{-}\Gamma_{d\mu}\Gamma_{11}\partial_{j}\theta_{-})(\partial_{k}\overline{\theta}_{+}\Gamma^{\mu}\Gamma^{\nu ab}\Gamma^{d}\theta_{-})+\nonumber \\
3(\overline{\theta}_{-}\Gamma^{\mu}\Gamma_{11}\partial_{k}\theta_{+})(\partial_{j}\overline{\theta}_{-}\Gamma_{\mu}\Gamma^{\nu ab}\theta_{-})+
3(\overline{\theta}_{-}\partial_{j}\theta_{-})(\partial_{k}\overline{\theta}_{+}\Gamma_{11}\Gamma^{\nu ab}\theta_{-})-\nonumber \\
3(\overline{\theta}_{-}\Gamma_{11}\Gamma^{d}\partial_{j}\theta_{-})(\overline{\theta}_{-}\Gamma^{\nu ab}\Gamma_{d}\partial_{k}\theta_{-})]\end{eqnarray}

\begin{eqnarray}
\overline{F}^{lmn}=\frac{1}{48}T_{NR}\epsilon_{0jk}[-\frac{1}{4}(\overline{\theta}_{+}\Gamma^{d}\partial_{k}\theta_{-})(\overline{\theta}_{-}\Gamma^{lmn}
\Gamma_{d}\partial_{j}\theta_{-}+\overline{\theta}_{-}\Gamma_{d}\Gamma^{lmn}\partial_{j}\theta_{-})-\nonumber \\
\frac{1}{32}(\overline{\theta}_{-}\Gamma^{d}\partial_{k}\theta_{+})(5(\overline{\theta}_{-}\Gamma^{lmn}\Gamma_{d}\partial_{j}\theta_{-})+6(\overline{\theta}
_{-}\Gamma_{d}\Gamma^{lmn}\partial_{j}\theta_{-}))+\frac{1}{32}(\overline{\theta}_{-}\Gamma_{\mu}\partial_{j}\theta_{-})\nonumber \\
(\partial_{k}\overline{\theta}_{+}\Gamma^{\mu lmn}\theta_{-})+\frac{1}{32}(\overline{\theta}_{-}\Gamma_{\mu d}\partial_{j}\theta_{+})(\overline{\theta}_{-}\Gamma^{d}\Gamma^{\mu lmn}\partial_{k}\theta_{-})+\frac{1}{32}(\partial_{k}\overline{\theta}_{+}\Gamma_{11}\Gamma_{\mu}\theta_{-})\nonumber \\
(\partial_{j}\overline{\theta}_{-}\Gamma_{11}\Gamma^{\mu lmn}\theta_{-})-\frac{1}{32}(\overline{\theta}_{-}\Gamma_{11}\Gamma_{d}\partial_{k}\theta_{-})(\overline{\theta}_{-}\Gamma_{11}\Gamma^{lmnd}\partial_{j}
\theta_{+})+\frac{1}{32}(\overline{\theta}_{-}\Gamma_{11}\partial_{k}\theta_{+})\nonumber \\
(\overline{\theta}_{-}\Gamma^{lmn}\Gamma_{11}\partial_{j}\theta_{-})-\frac{1}{32}(\partial_{k}\overline{\theta}_{+}\Gamma^{df}\Gamma_{11}\theta_{-})
(\partial_{j}\overline{\theta}_{-}\Gamma_{11}\Gamma_{d}\Gamma^{lmn}\Gamma_{f}\theta_{-})+\frac{1}{16}(\partial_{j}\overline{\theta}_{-}\Gamma^{df}\Gamma_{11}
\theta_{+})\nonumber \\
(\partial_{k}\overline{\theta}_{-}\Gamma_{11}\Gamma_{d}\Gamma^{lmn}\Gamma_{f}\theta_{-})-\frac{3}{32}(\overline{\theta}_{-}\Gamma_{11}\Gamma^{l}
\partial_{k}\theta_{-})(\overline{\theta}_{-}\Gamma_{11}\Gamma^{mn}\partial_{j}\theta_{+}-4(\overline{\theta}_{+}\Gamma_{11}\Gamma^{mn}\partial_{j}\theta_{-}))
-\nonumber  \\
\frac{1}{32}((\overline{\theta}_{-}\Gamma_{11}\Gamma^{d\mu}\partial_{j}\theta_{-})(\partial_{k}\overline{\theta}_{+}\Gamma_{11}\Gamma^{lmn}\Gamma_{d\mu}\theta_{-})+
(\overline{\theta}_{-}\Gamma_{df}\partial_{j}\theta_{-})(\partial_{k}\overline{\theta}_{+}\Gamma^{d}\Gamma^{lmn}\Gamma^{f}\theta_{-})+\nonumber \\
(\overline{\theta}_{-}\partial_{j}\theta_{-})(\partial_{k}\overline{\theta}_{+}\Gamma^{lmn}\theta_{-}))-\frac{1}{16}(\overline{\theta}_{-}
\Gamma^{df}\partial_{j}\theta_{-})(\overline{\theta}_{+}\Gamma_{d}\Gamma^{lmn}\Gamma_{f}\partial_{k}\theta_{-})]-\nonumber \\
\frac{1}{96\times4!\times
5!}\epsilon_{defglmn}\delta^{[abcd]}_{defg}[-5(\overline{\theta}_{-}\partial_{j}\theta_{-})(\partial_{k}\overline{\theta}_{+}\Gamma_{11}
\Gamma^{abcd}\theta_{-})+3(\partial_{j}\overline{\theta}_{-}\Gamma_{\mu}\theta_{-})\nonumber \\
(\overline{\theta}_{-}\Gamma_{11}\Gamma^{\mu
abcd}\partial_{k}\theta_{+})+(\partial_{j}\theta_{-}\Gamma_{11}\Gamma^{f}\theta_{-})(7(\overline{\theta}_{-}\Gamma^{abcd}\Gamma_{f}\partial_{k}\theta_{+})-
2(\overline{\theta}_{+}(\Gamma^{abcd}\Gamma_{f}+\nonumber \\
\Gamma_{f}\Gamma^{abcd})\partial_{k}\theta_{-}))-(\partial_{k}\overline{\theta}_{+}\Gamma_{f}\theta_{-})(\overline{\theta}_{-}\Gamma^{abcd}\Gamma_{11}
\Gamma^{f}\partial_{j}\theta_{-}+6(\overline{\theta}_{-}\Gamma^{f}\Gamma^{abcd}\Gamma_{11}\partial_{j}\theta_{-}))-\nonumber
\\
 8(\partial_{k}\overline{\theta}_{-}\Gamma_{f}\theta_{+})(\overline{\theta}_{-}(\Gamma^{abcd}\Gamma^{f}+\Gamma^{f}\Gamma^{abcd})\Gamma_{11}\partial_{j}
 \theta_{-})+(\overline{\theta}_{-}\Gamma_{\mu
 f}\partial_{j}\theta_{+})(\partial_{k}\overline{\theta}_{-}\Gamma_{11}\Gamma^{abcd}\Gamma^{\mu
 f}\theta_{-}+\nonumber \\
 2(\partial_{k}\overline{\theta}_{-}\Gamma^{\mu
 f}\Gamma^{abcd}\Gamma_{11}\theta_{-}))-(\overline{\theta}_{-}\Gamma_{11}\Gamma_{\mu
 f}\partial_{j}\theta_{-})(\partial_{k}\overline{\theta}_{+}\Gamma^{abcd}\Gamma^{\mu
 f}\theta_{-}+2(\partial_{k}\overline{\theta}_{+}\Gamma^{\mu
 f}\Gamma^{abcd}\theta_{-}))+\nonumber \\
 (\partial_{j}\overline{\theta}_{-}\Gamma^{f}\Gamma^{abcd}\Gamma^{e}\theta_{-})(\partial_{k}\overline{\theta}_{+}\Gamma_{fe}\Gamma_{11}\theta_{-}+
 2(\partial_{k}\overline{\theta}_{-}\Gamma_{fe}\Gamma_{11}\theta_{+}))-5(\overline{\theta}_{-}\Gamma^{abcd}\partial_{j}\theta_{-})\nonumber \\
 (\overline{\theta}_{-}\Gamma_{11}\partial_{k}\theta_{+})-(\overline{\theta}_{-}\Gamma_{ef}\partial_{k}\theta_{-})(\overline{\theta}_{-}
 \Gamma^{e}\Gamma^{abcd}\Gamma^{f}\partial_{j}\theta_{+}+2(\overline{\theta}_{+}\Gamma^{e}\Gamma^{abcd}\Gamma^{f}\Gamma_{11}\partial_{j}\theta_{-}))+
 \nonumber \\
 3(\partial_{j}\overline{\theta}_{-}\Gamma^{\mu
 abcd}\theta_{-})(\partial_{k}\overline{\theta}_{+}\Gamma_{11}\Gamma_{\mu}\theta_{-})]
 \end{eqnarray}

\begin{eqnarray}
\overline{F}_{\nu
abc}=\frac{1}{288}T_{NR}\epsilon_{0jk}[(\overline{\theta}_{-}\Gamma^{\mu}\partial_{j}\theta_{-})(3(\overline{\theta}_{-}\Gamma_{\mu\nu
abc
}\partial_{k}\theta_{+})-23(\overline{\theta}_{-}\Gamma_{abc}\partial_{k}\theta_{+})\eta_{\mu\nu})+\nonumber
\\
(\overline{\theta}_{-}\Gamma^{d}\partial_{j}\theta_{+})(5(\overline{\theta}_{-}\Gamma_{\nu
abcd }\partial_{k}\theta_{-})+3(\overline{\theta}_{-}\Gamma_{\nu
bc}\partial_{k}\theta_{-})\eta_{ad})-24(\overline{\theta}_{+}\Gamma_{a}\partial_{j}\theta_{-})\nonumber
\\
(\overline{\theta}_{-}\Gamma_{\nu
bc}\partial_{k}\theta_{-})+(\overline{\theta}_{-}\Gamma_{11}\Gamma^{\mu}\partial_{k}\theta_{+})(-3(\overline{\theta}_{-}\Gamma_{11}\Gamma_{abc}\partial_{j}
\theta_{-})\eta_{\mu\nu}+5(\partial_{j}\overline{\theta}_{-}\Gamma_{11}\Gamma_{\mu\nu
abc }\theta_{-}))+\nonumber \\
(\overline{\theta}_{-}\Gamma_{11}\Gamma^{d}\partial_{j}\theta_{-})(\overline{\theta}_{-}\Gamma_{11}\Gamma_{\nu
abc
}\Gamma_{d}\partial_{k}\theta_{+}-4(\overline{\theta}_{-}\Gamma_{d}\Gamma_{\nu
abc }\Gamma_{11}\partial_{k}\theta_{+})+\nonumber \\
12(\overline{\theta}_{+}\Gamma_{\nu bc
}\Gamma_{11}\partial_{k}\theta_{-})\eta_{ad})+(\overline{\theta}_{-}\Gamma^{\mu
d}\partial_{k}\theta_{+})(\partial_{j}\overline{\theta}_{-}\Gamma_{\mu}\Gamma_{\nu
abc
}\Gamma_{d}\theta_{-}-4(\partial_{j}\overline{\theta}_{-}\Gamma_{d}\Gamma_{\nu
abc }\Gamma_{\mu}\theta_{-}))-\nonumber \\
(\overline{\theta}_{-}\Gamma^{df}\partial_{k}\theta_{-})(\overline{\theta}_{-}\Gamma_{d}\Gamma_{\nu
abc
}\Gamma_{f}\partial_{j}\theta_{+}+2(\overline{\theta}_{+}\Gamma_{d}\Gamma_{\nu
abc
}\Gamma_{f}\partial_{j}\theta_{-}))+(\partial_{k}\overline{\theta}_{-}\Gamma_{d}\Gamma_{11}\Gamma_{\nu
abc }\Gamma_{f}\theta_{-})\nonumber \\
(3(\partial_{j}\overline{\theta}_{+}\Gamma^{df}\Gamma_{11}\theta_{-})+2(\partial_{j}\overline{\theta}_{-}\Gamma^{df}\Gamma_{11}\theta_{+}))+5(
\partial_{j}\overline{\theta}_{+}\Gamma_{11}\theta_{-})(\partial_{k}\overline{\theta}_{-}\Gamma_{11}\Gamma_{\nu
abc}\theta_{-})+\nonumber \\
3(\overline{\theta}_{-}\Gamma_{11}\Gamma_{\mu
d}\partial_{j}\theta_{-})(\partial_{k}\overline{\theta}_{+}\Gamma^{\mu}\Gamma_{11}\Gamma_{\nu
abc
}\Gamma^{d}\theta_{-})-3(\overline{\theta}_{-}\partial_{j}\theta_{-})(\partial_{k}\overline{\theta}_{+}\Gamma_{\nu
abc }\theta_{-})]
\end{eqnarray}

\begin{eqnarray}
F^{\nu abdf}=\frac{1}{96}T_{NR}\epsilon_{0jk}[6(\overline{\theta}_{-}\Gamma^{c}\partial_{j}\theta_{+})(\overline{\theta}_{-}
\Gamma^{\nu
abdf}\Gamma_{c}\partial_{k}\theta_{+}-\overline{\theta}_{-}\Gamma_{c}\Gamma^{\nu
abdf }\partial_{k}\theta_{+})-\nonumber \\
2(\overline{\theta}_{-}\Gamma_{11}\Gamma_{\mu}\partial_{k}\theta_{+})(\partial_{j}\overline{\theta}_{+}\Gamma_{11}\Gamma^{\mu}\Gamma^{\nu
abdf
}\theta_{-}+\partial_{j}\overline{\theta}_{+}\Gamma_{11}\Gamma^{\nu
abdf
}\Gamma^{\mu}\theta_{-})-2(\overline{\theta}_{-}\Gamma_{11}\Gamma_{\rho\sigma}\partial_{k}\theta_{+})\nonumber
\\
(\partial_{j}\overline{\theta}_{+}\Gamma_{11}\Gamma^{\rho}\Gamma^{\nu
abdf
}\Gamma^{\sigma}\theta_{-})+2(\overline{\theta}_{-}\Gamma^{c}\partial_{j}\theta_{+})(\partial_{k}\overline{\theta}_{-}\Gamma^{\nu
adf
}\Gamma_{c}\theta_{+}+\partial_{k}\overline{\theta}_{-}\Gamma_{c}\Gamma^{\nu
abdf }\theta_{+})+\nonumber \\
3(\overline{\theta}_{+}\Gamma^{c}\partial_{j}\theta_{-})(-2(\partial_{k}\overline{\theta}_{+}\Gamma^{\nu
abdf }\Gamma_{c}\theta_{-})+\overline{\theta}_{-}\Gamma^{\nu
abdf}\Gamma_{c}\partial_{k}\theta_{+})-4(\overline{\theta}_{-}\Gamma_{11}\Gamma_{\mu}\partial_{k}\theta_{+})\nonumber
\\
(\partial_{j}\overline{\theta}_{-}\Gamma_{11}\Gamma^{\mu}\Gamma^{\nu
abdf
}\theta_{+})+(\overline{\theta}_{+}\Gamma_{11}\Gamma_{\mu}\partial_{k}\theta_{-})(\partial_{j}\overline{\theta}_{+}\Gamma_{11}\Gamma^{\mu}
\Gamma^{\nu abdf}\theta_{-}-\nonumber \\
2(\partial_{j}\overline{\theta}_{+}\Gamma_{11}\Gamma^{\nu abdf
}\Gamma^{\mu}\theta_{-}))-2(\partial_{j}\overline{\theta}_{-}\Gamma^{\mu}\theta_{-})(\partial_{k}\overline{\theta}_{+}\Gamma_{\mu}\Gamma^{\nu
abdf }\theta_{+})+2(\partial_{k}\overline{\theta}_{+}\Gamma_{\mu
c}\theta_{-})\nonumber \\
(\overline{\theta}_{+}\Gamma^{c}\Gamma^{\nu
abdf}\Gamma^{\mu}\partial_{j}\theta_{-})+2(\overline{\theta}_{-}\Gamma^{c}\Gamma_{11}\partial_{j}\theta_{-})(\partial_{k}\overline{\theta}_{+}\Gamma_{11}
\Gamma^{\nu abdf}\Gamma_{c}\theta_{+}-\nonumber \\
2(\partial_{k}\overline{\theta}_{+}\Gamma_{11}\Gamma_{c}\Gamma^{\nu
abdf
}\theta_{+}))+2(\partial_{k}\overline{\theta}_{+}\Gamma_{11}\theta_{-})(\overline{\theta}_{+}\Gamma_{11}\Gamma^{\nu
abdf
}\partial_{j}\theta_{-})+(\overline{\theta}_{+}\Gamma^{\mu}\partial_{k}\theta_{+})\nonumber
\\
 (6(\partial_{j}\overline{\theta}_{-}\Gamma_{\mu}\Gamma^{\nu abdf
}\theta_{-})+5(\partial_{j}\theta_{-}\Gamma^{\nu
abdf}\Gamma_{\mu}\theta_{-}))+7(\overline{\theta}_{+}\Gamma_{\mu\sigma}\partial_{k}\theta_{+})(\partial_{j}\overline{\theta}_{-}\Gamma^{\sigma}\Gamma^{\nu abdf}\Gamma^{\mu}\theta_{-})-\nonumber  \\
 3(\overline{\theta}_{+}\Gamma_{\mu c}\partial_{j}\theta_{-})(\partial_{k}\overline{\theta}_{+}\Gamma^{c}\Gamma^{\nu abdf}\Gamma^{\mu}\theta_{-})
-4(\overline{\theta}_{+}\Gamma_{11}\partial_{j}\theta_{-})(\partial_{k}\overline{\theta}_{+}\Gamma_{11}\Gamma^{\nu abdf}\theta_{-})+\nonumber \\
(\partial_{k}\overline{\theta}_{+}\Gamma_{\mu c}\Gamma_{11}\theta_{+})(\partial_{j}\overline{\theta}_{-}\Gamma^{c}\Gamma_{11}\Gamma^{\nu abdf}\Gamma^{\mu}\theta_{-})+(\partial_{k}\overline{\theta}_{+}\Gamma_{\mu}\Gamma_{\sigma}\theta_{+})\nonumber \\
(\partial_{j}\overline{\theta}_{-}\Gamma^{\sigma}\Gamma^{\nu abdf}\Gamma^{\mu}\theta_{-})+(\overline{\theta}_{+}\Gamma_{11}\Gamma_{\sigma}\Gamma_{\mu}\partial_{j}\theta_{-})(\partial_{k}\overline{\theta}_{+}
\Gamma^{\sigma}\Gamma_{11}\Gamma^{\nu abdf}\Gamma^{\mu}\theta_{-})-\nonumber \\
(\partial_{k}\overline{\theta}_{+}\Gamma_{11}\Gamma_{c}\theta_{+})(\partial_{j}\overline{\theta}_{-}\Gamma_{11}\Gamma^{c}\Gamma^{\nu abdf}\theta_{-})+2(\overline{\theta}_{-}\Gamma_{ce}\partial_{k}\theta_{-})(\overline{\theta}_{+}\Gamma^{c}\Gamma^{\nu abdf}\Gamma^{e}\partial_{j}\theta_{+})-\nonumber \\
\frac{1}{4!}\epsilon_{cgeabdf}\delta^{[lmn]}_{cge}\frac{T_{NR}}{576}[-36(\overline{\theta}_{-}\Gamma^{l}\partial_{j}\theta_{+})(\partial_{k}\overline{\theta}
_{+}\Gamma^{\nu mn}\Gamma_{11}\theta_{-})+2(\overline{\theta}_{-}\Gamma^{r}\partial_{j}\theta_{+})\nonumber \\
(\partial_{k}\overline{\theta}_{-}\Gamma_{11}\Gamma^{\nu lmn}\Gamma_{r}\theta_{+}+\partial_{k}\overline{\theta}_{-}\Gamma_{11}\Gamma_{r}\Gamma^{\nu lmn}\theta_{+})+3(\overline{\theta}_{+}\Gamma^{r}\partial_{j}\theta_{-})\nonumber \\
(\partial_{k}\overline{\theta}_{+}\Gamma_{r}\Gamma^{\nu lmn}\Gamma_{11}\theta_{-}+2(\partial_{k}\overline{\theta}_{+}\Gamma_{11}\Gamma^{\nu lmn}\Gamma_{r}\theta_{-}))-4(\overline{\theta}_{-}\Gamma_{11}\Gamma_{\mu}\partial_{k}\theta_{+})\nonumber \\
(\partial_{j}\overline{\theta}_{-}\Gamma^{\mu}\Gamma^{\nu lmn}\theta_{+})-(\overline{\theta}_{+}\Gamma_{11}\Gamma_{\mu}\partial_{k}\theta_{-})
(2(\partial_{j}\overline{\theta}_{+}\Gamma^{\nu lmn}\Gamma^{\mu}\theta_{-})+\partial_{j}\overline{\theta}_{+}\Gamma^{\mu}\Gamma^{\nu lmn}\theta_{-})+\nonumber \\
2(\overline{\theta}_{-}\Gamma^{\mu}\partial_{j}\theta_{-})(\partial_{k}\overline{\theta}_{+}\Gamma_{11}\Gamma_{\mu}\Gamma^{\nu lmn}\theta_{+})-2(
\partial_{k}\overline{\theta}_{+}\Gamma_{r\mu}\theta_{-})(\overline{\theta}_{+}\Gamma^{r}\Gamma^{\nu lmn}\Gamma_{11}\Gamma^{\mu}\partial_{j}\theta_{-})+\nonumber \\
2(\overline{\theta}_{-}\Gamma^{r}\Gamma_{11}\partial_{j}\theta_{-})(-2(\partial_{k}\overline{\theta}_{+}\Gamma_{r}\Gamma^{\nu lmn}\theta_{+})+
\partial_{k}\overline{\theta}_{+}\Gamma^{\nu lmn}\Gamma_{r}\theta_{+})+2(\partial_{k}\overline{\theta}_{+}\Gamma_{11}\theta_{-})\nonumber \\
(\overline{\theta}_{+}\Gamma^{\nu lmn}\partial_{j}\theta_{-})+6(\overline{\theta}_{+}\Gamma_{\sigma\mu}\partial_{k}\theta_{+})(\partial_{j}\overline{\theta}_{-}\Gamma^{\mu}\Gamma^{\nu lmn}\Gamma_{11}\Gamma^{\sigma}\theta_{-})-(\partial_{k}\overline{\theta}_{+}\theta_{+})\nonumber \\
(\partial_{j}\overline{\theta}_{-}\Gamma^{\nu lmn}\Gamma_{11}\theta_{-})+(\partial_{k}\overline{\theta}_{+}\Gamma_{\mu r}\Gamma_{11}\theta_{+})(
\partial_{j}\overline{\theta}_{-}\Gamma^{r}\Gamma^{\nu lmn}\Gamma^{\mu}\theta_{-})+(\partial_{j}\overline{\theta}_{-}\Gamma_{11}\Gamma_{\mu\sigma}\theta_{+})\nonumber  \\
(\partial_{k}\overline{\theta}_{+}\Gamma^{\mu}\Gamma^{\nu lmn}\Gamma^{\sigma}\theta_{-})+3(\partial_{j}\overline{\theta}_{-}\Gamma_{11}\theta_{+})(\partial_{k}\overline{\theta}_{+}\Gamma^{\nu lmn}\theta_{-})+(\partial_{k}\overline{\theta}_{+}\Gamma_{11}\Gamma^{r}\theta_{+})\nonumber \\
(\partial_{j}\overline{\theta}_{-}\Gamma_{r}\Gamma^{\nu lmn}\theta_{-})+3(\overline{\theta}_{+}\Gamma_{r\sigma}\partial_{j}\theta_{-})(\partial_{k}\overline{\theta}_{+}\Gamma^{r}\Gamma^{\nu lmn}\Gamma_{11}\Gamma^{\sigma}\theta_{-})+(\overline{\theta}_{+}\Gamma^{\mu}\partial_{k}\theta_{+})\nonumber \\
(6(\partial_{j}\overline{\theta}_{-}\Gamma_{\mu}\Gamma^{\nu lmn}\Gamma_{11}\theta_{-})+5(\partial_{j}\overline{\theta}_{-}\Gamma^{\nu lmn}\Gamma_{11}\Gamma_{\mu}\theta_{-}))+2(\overline{\theta}_{-}\Gamma_{rs}\partial_{k}\theta_{-})\nonumber \\
(\overline{\theta}_{+}\Gamma^{r}\Gamma^{\nu lmn}\Gamma_{11}\Gamma^{s}\partial_{j}\theta_{+})]\end{eqnarray}

\begin{eqnarray}
\overline{F}_{fglmn}=\frac{1}{96\times
5!}T_{NR}\epsilon_{0jk}[3(\overline{\theta}_{-}\Gamma^{\mu}\partial_{j}\theta_{-})(-5(\overline{\theta}_{-}\Gamma_{\mu
fglmn }\partial_{k}\theta_{+})+\nonumber \\
4(\overline{\theta}_{+}\Gamma_{\mu
fglmn}\partial_{k}\theta_{-}))+(\overline{\theta}_{-}\Gamma^{a}\partial_{j}\theta_{+})(17(\overline{\theta}_{-}\Gamma_{fglmn}\Gamma_{a}\partial
_{k}\theta_{-})-22(\overline{\theta}_{-}\Gamma_{a}\Gamma_{fglmn}\partial_{k}\theta_{-}))+\nonumber
\\
4(\overline{\theta}_{+}\Gamma^{a}\partial_{j}\theta_{-})(\overline{\theta}_{-}\Gamma_{fglmn}\Gamma_{a}\partial_{k}\theta_{-}-5(\overline{\theta}_{-}
\Gamma_{a}\Gamma_{fglmn}\partial_{k}\theta_{-}))+9(\overline{\theta}_{-}\Gamma_{11}\Gamma^{\mu}\partial_{k}\theta_{+})\nonumber
\\
(\partial_{j}\overline{\theta}_{-}\Gamma_{11}\Gamma_{\mu
fglmn}\theta_{-})+12(\overline{\theta}_{+}\Gamma_{11}\Gamma^{\mu}\partial_{k}\theta_{-})(\partial_{j}\overline{\theta}_{-}\Gamma_{\mu
fglmn
}\Gamma_{11}\theta_{-})+(\partial_{j}\overline{\theta}_{-}\Gamma_{11}\Gamma^{a}\theta_{-})\nonumber
\\
(-4(\partial_{k}\overline{\theta}_{+}\Gamma_{11}\Gamma_{fglmn}\Gamma_{a}\theta_{-})+13(\partial_{k}\overline{\theta}_{+}\Gamma_{11}\Gamma_{a}
\Gamma_{fglmn}\theta_{-})+6(\partial_{k}\overline{\theta}_{-}\Gamma_{11}\Gamma_{a}\Gamma_{fglmn}\theta_{+})+\nonumber
\\
2(\partial_{k}\overline{\theta}_{-}\Gamma_{11}\Gamma_{fglmn}\Gamma_{a}\theta_{+}))-(\partial_{k}\overline{\theta}_{-}\Gamma_{11}\Gamma^{a}
\Gamma_{fglmn}\Gamma^{b}\theta_{-})(4(\partial_{j}\overline{\theta}_{-}\Gamma_{b}\Gamma_{a}\Gamma_{11}\theta_{+})+\nonumber
\\
3(\partial_{j}\overline{\theta}_{+}\Gamma_{b}\Gamma_{a}\Gamma_{11}\theta_{-}))+3(\overline{\theta}_{-}\Gamma_{b}\Gamma_{a}\partial_{j}\theta_{-}
)(\partial_{k}\overline{\theta}_{+}\Gamma^{b}\Gamma_{fglmn}\Gamma^{a}\theta_{-})+2(\overline{\theta}_{-}\Gamma^{ab}\partial_{k}\theta_{-})\nonumber
\\
(\overline{\theta}_{-}\Gamma_{a}\Gamma_{fglmn}\Gamma_{b}\partial_{j}\theta_{+}+\overline{\theta}_{+}\Gamma_{a}\Gamma_{fglmn}\Gamma_{b}\partial_{j}
\theta_{-})-6(\overline{\theta}_{-}\Gamma_{11}\partial_{k}\theta_{+})(\partial_{j}\overline{\theta}_{-}\Gamma_{11}\Gamma_{fglmn}\theta_{-})+\nonumber
\\
12(\partial_{j}\overline{\theta}_{-}\Gamma_{11}\theta_{+})(\partial_{k}\overline{\theta}_{-}\Gamma_{11}\Gamma_{fglmn}\theta_{-})-3(\partial_{k}
\overline{\theta}_{+}\Gamma_{11}\Gamma_{fglmn}\Gamma^{\mu
a}\theta_{-})(\overline{\theta}_{-}\Gamma_{11}\Gamma_{\mu
a}\partial_{j}\theta_{-})-\nonumber \\
(\overline{\theta}_{-}\Gamma^{\mu
a}\partial_{k}\theta_{+})(4(\partial_{j}\overline{\theta}_{-}\Gamma_{\mu
a
}\Gamma_{fglmn}\theta_{-})-\partial_{j}\overline{\theta}_{-}\Gamma_{fglmn}\Gamma_{\mu
a}\theta_{-})-2(\overline{\theta}_{+}\Gamma^{\mu
a}\partial_{k}\theta_{-})\nonumber \\
(\overline{\theta}_{-}\Gamma_{\mu
a}\Gamma_{fglmn}\partial_{j}\theta_{-}+\partial_{j}\Gamma_{\mu
a}\Gamma_{fglmn}\theta_{-})]-\nonumber \\
\frac{1}{192\times
5!}\epsilon_{0jk}\epsilon_{defglmn}\delta^{ba}_{de}[(16(\overline{\theta}_{+}\Gamma_{c}\partial_{j}\theta_{-})
+21(\overline{\theta}_{-}\Gamma_{c}\partial_{j}\theta_{+}))(\overline{\theta}_{-}\Gamma^{abc}\Gamma_{11}\partial_{k}\theta_{-})+\nonumber
\\
8(7(\overline{\theta}_{+}\Gamma^{a}\partial_{j}\theta_{-})+3(\overline{\theta}_{-}\Gamma^{a}\partial_{j}\theta_{+}))(\overline{\theta}_{-}
\Gamma^{b}\Gamma_{11}\partial_{k}\theta_{-})-(\overline{\theta}_{-}\Gamma_{\mu}\partial_{j}\theta_{-})\nonumber
\\
 (12(\overline{\theta}_{+}\Gamma^{\mu
 ab}\Gamma_{11}\partial_{k}\theta_{-})+11(\overline{\theta}_{-}\Gamma_{11}\Gamma^{\mu
 ab}\partial_{k}\theta_{+}))+8(\overline{\theta}_{-}\Gamma_{\mu}\Gamma^{b}\Gamma_{11}\partial_{j}\theta_{-})(\overline{\theta}_{+}\Gamma^{\mu a}
 \partial_{k}\theta_{-}+\nonumber \\
 \overline{\theta}_{-}\Gamma^{\mu a}\partial_{k}\theta_{+})+(\partial_{k}\overline{\theta}_{+}\Gamma_{\mu
 c}\theta_{-})(\partial_{j}\overline{\theta}_{-}\Gamma^{\mu
 abc}\Gamma_{11}\theta_{-}+2(\overline{\theta}_{-}\Gamma^{\mu
 b}\Gamma_{11}\partial_{j}\theta_{-})\eta^{ac})+\nonumber \\
 (\overline{\theta}_{-}\Gamma_{11}\Gamma_{c}\partial_{k}\theta_{-})(-8(\overline{\theta}_{+}\Gamma^{abc}\partial_{j}\theta_{-})+3(\overline{\theta}_{-}
 \Gamma^{abc}\partial_{j}\theta_{+}))+(\overline{\theta}_{-}\Gamma^{\mu
 ab}\partial_{j}\theta_{-})\nonumber \\
 (12(\overline{\theta}_{+}\Gamma_{11}\Gamma_{\mu}\partial_{k}\theta_{-})-17(\overline{\theta}_{-}\Gamma_{11}\Gamma_{\mu}
 \partial_{k}\theta_{+}))-4(\partial_{j}\overline{\theta}_{-}\Gamma_{cd}\Gamma_{11}\theta_{+})(\partial_{k}\overline{\theta}_{-}\Gamma^{c}\Gamma^{ab}
 \Gamma^{d}\theta_{-})+\nonumber \\
 (\partial_{j}\overline{\theta}_{+}\Gamma_{cd}\Gamma_{11}\theta_{-})(2(\partial_{k}\overline{\theta}_{-}\Gamma^{cd}\Gamma^{ab}\theta_{-})-17(
 \overline{\theta}_{-}\partial_{k}\theta_{-})\eta^{ac}\eta^{bd})+(\overline{\theta}_{-}\Gamma_{cd}\partial_{k}\theta_{-})\nonumber \\
 (11(\partial_{j}\overline{\theta}_{+}\Gamma_{11}\theta_{-})\eta^{ac}\eta^{bd}+2(-\partial_{j}\overline{\theta}_{+}\Gamma^{cd}\Gamma^{ab}\Gamma_{11}
 \theta_{-}+\overline{\theta}_{-}\Gamma^{c}\Gamma^{ab}\Gamma^{d}\Gamma_{11}\partial_{j}\theta_{+}-\nonumber \\
 \overline{\theta}_{+}\Gamma^{c}\Gamma^{ab}
 \Gamma^{d}\Gamma_{11}\partial_{j}\theta_{-}))+3(\overline{\theta}_{-}\Gamma_{\mu
 c}\Gamma_{11}\partial_{j}\theta_{-})(\partial_{k}\overline{\theta}_{+}\Gamma^{\mu abc}\theta_{-}+2(\partial_{k}\overline{\theta}_{+}\Gamma^{\mu
 b}\theta_{-})\eta^{ac})]
\end{eqnarray}

\newpage


\end{document}